\documentclass[10pt]{iopart}

\usepackage{iopams}  

\usepackage[superscript,biblabel]{cite}
\usepackage
[colorlinks]{hyperref}\hypersetup{
     colorlinks   = true,
     citecolor    = blue
}
\usepackage{hyperref}
\hypersetup{
    colorlinks=true,
    linkcolor=blue,
    filecolor=magenta,      
    urlcolor=cyan,
}
\usepackage[normalem]{ulem}
\usepackage[toc,page]{appendix}
\usepackage[utf8]{inputenc}
\usepackage[T1]{fontenc}
\usepackage{lmodern}
\usepackage[english]{babel}
\usepackage{color}
\usepackage{graphics}
\usepackage{epsfig}
\usepackage{float}
\usepackage{array}
\usepackage{bm}
\usepackage{footnote}
\usepackage{gensymb}
\usepackage[symbol]{footmisc}
\usepackage{array}
\usepackage{makecell}
\usepackage{tabularx}
\usepackage{pdfpages}
\newcolumntype{C}[1]{>{\centering\arraybackslash$}p{#1}<{$}}
\usepackage{fancyhdr}
\usepackage{lastpage}

\usepackage[heightrounded]{geometry}
\usepackage{iopams} 

\begin{document}

\setlength{\headwidth}{\textwidth}
\pagestyle{fancyplain}
\lhead{ \fancyplain{}{Nucl. Fusion XX (2025) XXXXXX} }
\rhead{ \fancyplain{}{A. Kumar \it et  al} }
\setlength{\headwidth}{\textwidth}
\addtolength{\headwidth}{\marginparsep}
\addtolength{\headwidth}{\marginparwidth}

\title[]{Understanding Carbon Sourcing and Transport Originating from the Helicon Antenna Surfaces During High-Power Helicon Discharge in DIII-D Tokamak\mbox{*}}

\author{
A. Kumar$^{1\dagger}$, D. Nath$^{2}$, W. Tierens$^{1}$, J.D. Lore$^{1}$, 
R. Wilcox$^{1}$, G. Ronchi$^{1}$, M. Shafer$^{1}$, A. Y. Joshi$^{2}$, 
O. Sahni$^{2}$, M. S. Shephard$^{2}$, B. Van Compernolle$^{3}$, 
R.I. Pinsker$^{3}$, A. Demby$^{4}$, and O. Schmitz$^{4}$%
}

\address{$^{1}$Oak Ridge National Laboratory, 1 Bethel Valley Road, Oak Ridge, TN 37831, USA}
\address{$^{2}$Rensselaer Polytechnic Institute, Troy, NY 12180, USA}
\address{$^{3}$General Atomics, San Diego, CA 92186, USA}
\address{$^{4}$University of Wisconsin–Madison, Madison, WI 53706, USA}

\ead{$^\dagger$kumara@ornl.gov}


\begin{abstract}
The high-power helicon wave system in the DIII-D tokamak introduces new plasma--material interaction (PMI) challenges due to rectified RF sheath potentials forming near antenna structures and surrounding tiles. Using the STRIPE modeling framework---which integrates SOLPS-ITER, COMSOL, RustBCA, and GITR/GITRm---we simulate carbon erosion, re-deposition, and global impurity transport in two H-mode discharges with varying antenna--plasma gaps and RF powers. COMSOL predicts rectified sheath potentials of 1--5~kV, localized near the bottom of the antenna where magnetic field lines intersect at grazing angles. Erosion is dominated by carbon self-sputtering, with RF-accelerated D$^+$ ions contributing up to 1\% of the total erosion flux. GITRm simulations show that in the small-gap case, only $\sim$13\% of eroded carbon is re-deposited locally, with 58\% transported into the core. In contrast, the large-gap case exhibits lower total erosion, along with reduced core penetration ($\sim$35\%) and weaker re-deposition ($\sim$4\%), consistent with lower collisionality and limited plasma contact. The simulation trends are consistent with experimental observations, which have not shown elevated core impurity levels during helicon operation in the present graphite-wall configuration. However, under certain plasma conditions and magnetic configurations, the helicon antenna may still act as a finite source of net erosion and core-directed impurity transport, potentially influencing the overall core impurity balance. These findings emphasize the need for sheath-aware antenna designs and predictive impurity transport modeling to support future high-power RF systems with high-$Z$ first wall materials in fusion devices.

\footnotetext{This manuscript has been authored by UT-Battelle, LLC, under contract DE-AC05-00OR22725 with the US Department of Energy (DOE). The US government retains and the publisher, by accepting the article for publication, acknowledges that the US government retains a nonexclusive, paid-up, irrevocable, worldwide license to publish or reproduce the published form of this manuscript, or allow others to do so, for US government purposes. DOE will provide public access to these results of federally sponsored research in accordance with the DOE Public Access Plan (http://energy.gov/downloads/doe-public-access-plan).}
\end{abstract}
%
%
%
%
\ioptwocol
\section{Introduction: Context and Objectives}
\label{sec:1}

Radio-frequency (RF) waves play a central role in magnetic confinement fusion (MCF) systems, providing critical capabilities for plasma heating, non-inductive current drive, and profile control. The operational value of RF-based heating was prominently demonstrated during the 2022 deuterium-tritium campaign at JET, where ion cyclotron resonance heating (ICRH) delivered approximately 20\% of the total 59~MJ fusion energy yield, contributing substantially to record plasma performance~\cite{Maslov_2023}. As the fusion community progresses toward compact pilot plants and steady-state reactor designs, exemplified by conceptual studies for devices such as SPARC and advanced tokamaks targeting net electricity production~\cite{Buttery:2021, Creely_2020, Rodriguez-Fernandez_2022}, efficient RF current drive is increasingly recognized as essential for sustaining high bootstrap fractions and avoiding reliance on inductive solenoidal operation.

Among RF techniques, helicon waves -- fast wave branch in the lower hybrid frequency range -- have re-emerged as a promising candidate for mid-radius current drive. In a recent  pilot plant concept study~\cite{Buttery:2021}, helicon current drive is noted for its potential to efficiently deliver current at mid-plasma radius while minimizing wave accessibility and damping constraints that limit lower hybrid current drive in high-density, high-field devices. Theoretical analyses have long predicted efficient helicon absorption via electron Landau damping and mode conversion, offering the possibility of fully non-inductive operation with favorable control of safety factor profiles and confinement stability~\cite{Theilhaber:1980}. However, early experimental efforts to couple helicon waves in tokamaks faced persistent challenges, including low antenna loading, weak core absorption, and practical engineering limitations~\cite{Pinsker:1994}.

These barriers are recently overcome through the development and deployment of a novel 30-element traveling-wave ``comb-line'' helicon antenna on DIII-D~\cite{Pinsker:2024, Compernoll:2021}. This system demonstrated the first successful high-power helicon coupling into diverted tokamak plasmas, achieving over 0.6~MW of coupled RF power at 476~MHz. The experiments provided unambiguous evidence of single-pass electron heating consistent with AORSA and GENRAY predictions, and the antenna showed robust load resilience across a range of plasma conditions, addressing long-standing uncertainties about practical implementation in reactor environments~\cite{Pinsker:2024}. These advances represent a pivotal step toward establishing helicon technology as a viable tool for advanced tokamak operation and fusion pilot plants.

Nevertheless, the integration of high-power helicon systems introduces new PMI challenges analogous to, and in some respects exceeding, those observed in ICRH systems. RF sheath rectification at plasma-facing components (PFCs) near the antenna can generate substantial time-averaged DC sheath potentials reaching hundreds of volts which in turn accelerate ions from the scrape-off layer (SOL) toward material surfaces. This process can significantly enhance sputtering yields for both deuterium fuel ions and light impurities, leading to increased erosion of antenna structures and adjacent tiles. Although recent DIII-D helicon experiments have not shown strong evidence of RF-specific impurity influxes during L-mode operation~\cite{Pinsker:2024}, the transition to H-mode may significantly increase impurity retention in the core due to improved confinement and reduced edge transport. As a result, the PMI response under H-mode conditions--particularly in configurations with varying RF power and antenna-plasma gaps--remains largely unexplored and requires detailed investigation.

To address these challenges, we apply the STRIPE (Simulated Transport of RF Impurity Production and Emission) framework~\cite{kumar:2025, kumar_rfppc_2025}, a multi-physics modeling workflow developed to simulate PMI in RF environments. STRIPE integrates high-fidelity tools including SOLPS-ITER, COMSOL, RustBCA, GITR, and GITRm to quantify sheath rectification, material erosion, impurity production, and whole-device 3D impurity transport originating from the RF antenna and adjacent PFCs. In this study, STRIPE is applied to analyze two DIII-D H-mode discharges (\#196154 and \#200882), combining experimentally constrained plasma parameters and sheath conditions to compute spatially resolved carbon erosion and impurity transport near the helicon antenna. Similar to previous STRIPE applications focused on tungsten erosion from ICRH antennas in WEST~\cite{kumar:2025, kumar_rfppc_2025}, this work concentrates on carbon erosion and its  transport originating from the newly installed helicon antenna in DIII-D tokamak. This regime is critical for assessing PMI risks in devices considering helicon current drive and for improving understanding of the coupling between RF sheath effects, impurity sourcing, and edge plasma conditions.

This paper is organized as follows: Section~\ref{sec:2} describes the STRIPE framework and its constituent modeling tools. Section~\ref{sec:3} outlines the helicon background plasma simulations performed with SOLPS-ITER for the DIII-D discharges. Section~\ref{sec:4} details the additional inputs and boundary conditions used in the STRIPE workflow. Section~\ref{sec:5} presents results on carbon erosion and global impurity transport under H-mode conditions. Section~\ref{sec:6} provides a detailed physics discussion of the results, including interpretation of erosion trends, sensitivity to plasma parameters, and limitations of the modeling approach. Finally, Section~\ref{sec:7} summarizes the main conclusions and outlines directions for future work.

\section{STRIPE - An Integrated Modeling Framework for RF-PMI}
\label{sec:2}

Accurate prediction of RF-induced PMI, including surface erosion and impurity transport, requires multi-scale modeling across sheath, surface, and plasma domains. STRIPE (Simulated Transport of RF Impurity Production and Emission) is a modular framework developed to integrate key physics processes driving PMI at RF antenna structures and adjacent PFCs.  The main inputs to this framework are:

\begin{itemize}
    \item \textbf{Plasma Background:} Electron density, temperature, and ion fluxes are provided by SOLPS-ITER simulations \cite{Bonnin:2016} of DIII-D H-mode discharges \#196154 and \#200882 with helicon operations, corresponding to larger and smaller antenna-plasma gaps, respectively. These profiles define plasma conditions at walls surrounding the helicon antenna (WSHA), enabling sheath and sputtering modeling. See Section~\ref{sec:3} for details.
    \item \textbf{Geometry Defeaturing:} Engineering CAD models of the helicon antenna are defeatured to retain relevant components   while improving meshing efficiency for finite-element simulations.
    
    \item \textbf{RF Sheaths:} Rectified sheath potentials are computed using the sheath-equivalent dielectric layer method \cite{Beers1, Beers2, Tierens_2024}. This approach models surface-adjacent layers with complex permittivity $\varepsilon$ and conductivity $ \sigma$, derived from local plasma parameters. Iterative COMSOL simulations yield spatially resolved 3D RF sheath potentials, which are used to model ion acceleration and surface energy-angle distributions.
    
    \item \textbf{Sputtering Yields:} The RustBCA code \cite{Drobny:2023, RustBCA} is used to calculate energy ($ \mathcal{E}$) and angle ($ \theta$) resolved sputtering yields, $ Y(\mathcal{E,\theta})$,  for deuterium   and carbon  atoms impacting carbon surfaces. 
    
    \item \textbf{Ion-Surface Interaction:} Ion trajectories from the sheath edge to the surface are tracked using the 3D Monte-Carlo particle tracer code-GITR\cite{Younkin:2021}, accounting for gyromotion, $ \mathbf{E} \times \mathbf{B}$ drifts, and sheath acceleration. Both thermal and RF sheath conditions are modeled using a common exponential electric field profile based on Brooks' model \cite{Brooks:1990}:
    
   \begin{eqnarray}
    E_n & = & V_{ sheath} \left[
    \frac{f_{ D}}{2\lambda_{ D}} \exp\left(-\frac{r}{2\lambda_{ D}}\right) \right. \nonumber \\
    & & \left. + \frac{1 - f_{ D}}{\rho_i} \exp\left(-\frac{r}{\rho_i}\right)
    \right]
    \label{Eq:RFsheath} 
    \end{eqnarray}

where $ V_{sheath}$ is the time-averaged rectified DC voltage arising on the antenna structures, \( r\) is the distance from the surface, \( \lambda_{ D}\) is the local Debye length, \(\rho_i\) is the local ion gyroradius, and \( f_{ D}\) is the fraction of the potential drop across the Debye sheath. $ V_{sheath}$ is either calculated from classical sheath theory for thermal cases or extracted from COMSOL for RF cases.

    \item \textbf{Erosion Rate Calculation:} At each surface element \(i\), the effective sputtering yield Y$_\mathrm{eff}$ is computed by integrating the product of incident ion energy-angle distributions (IEADs) $f_i(\mathcal{E},\theta)$ and corresponding RustBCA's  $ Y(\mathcal{E},\theta)$:
    
    \begin{equation}
    Y_{ eff}(i) = \int_{\theta=0}^{90^\circ} \int_{\mathcal{E}_{ min}}^{\mathcal{E}_{ max}} Y_i(\mathcal{E}, \theta) f_i(\mathcal{E}, \theta) \, d\theta \, d\mathcal{E}.
    \label{Eq:spYield}
    \end{equation}

    The gross carbon erosion flux is then given by:

    \begin{equation}
    \Gamma_{ gross} = \sum_{i=1}^{N} Y_{ eff}(i) \cdot \Gamma_{ ions}(i),
    \label{Eq:grossErosion}
    \end{equation}

    where \( \Gamma_{ ions}(i)\) is the ion flux from the SOLPS background.

    \item \textbf{Whole Device Impurity Transport:} Whole-device transport of C impurities originating from the surfaces surrounding the helicon antenna is simulated using the 3D unstructured-mesh Monte Carlo code GITRm~\cite{nath:2023, Nath:2025}. These simulations include modeling of net erosion, re-deposition, and C self-sputtering effects in the vicinity of the antenna and adjacent surfaces. Further details on the GITRm implementation, mesh generation for the helicon geometry, and simulation configuration are provided in \ref{app:gitrm}.

     \end{itemize}

 While the coupling between various computational tools within the STRIPE framework is currently performed in a sequential and modular fashion, the integrated framework enables comprehensive analysis of PMI and associated global impurity transport in RF environments of MCF experiments.
\begin{figure}
    \centering
    \includegraphics[width=\linewidth]{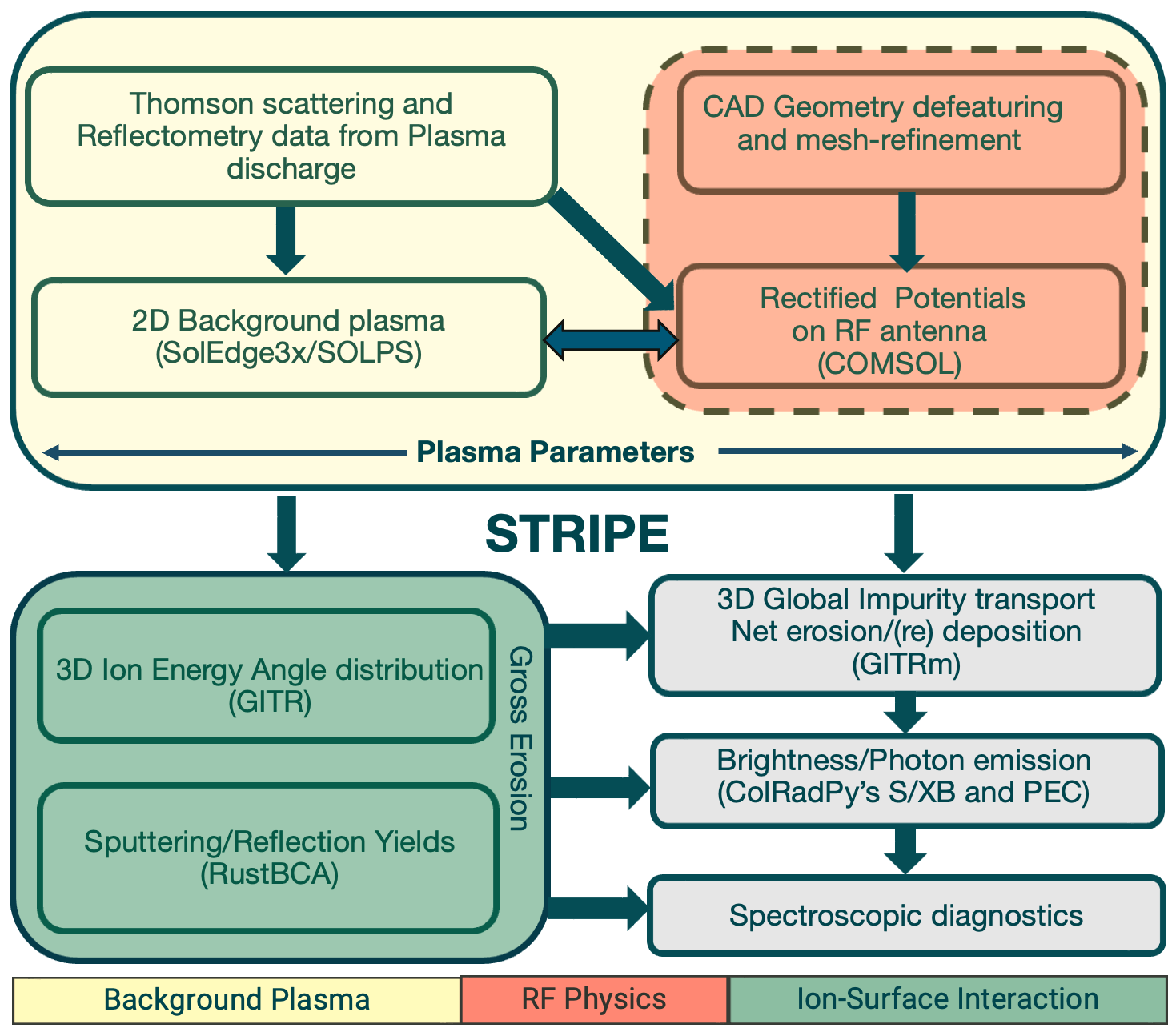}
    \caption{STRIPE workflow for modeling RF-induced erosion and impurity transport originating from RF antenna structures, coupling SOLPS/SolEdge3x, COMSOL, RustBCA, GITR/GITRm, and ColRadPy.}
    \label{fig:1}
\end{figure}

\section{Modeling of the DIII-D Background Plasma}
\label{sec:3}
\subsection{Discharge Summary }

This study focuses on two DIII-D H-mode discharges, \#196154 and \#200882, selected to assess the impact of antenna-plasma proximity on RF sheath formation, material erosion, impurity sourcing, and global transport. Both discharges operated with the DIII-D helicon antenna at 476~MHz, but differed significantly in edge magnetic geometry and coupled RF power.

\begin{figure}
    \centering
    \includegraphics[width=\linewidth]{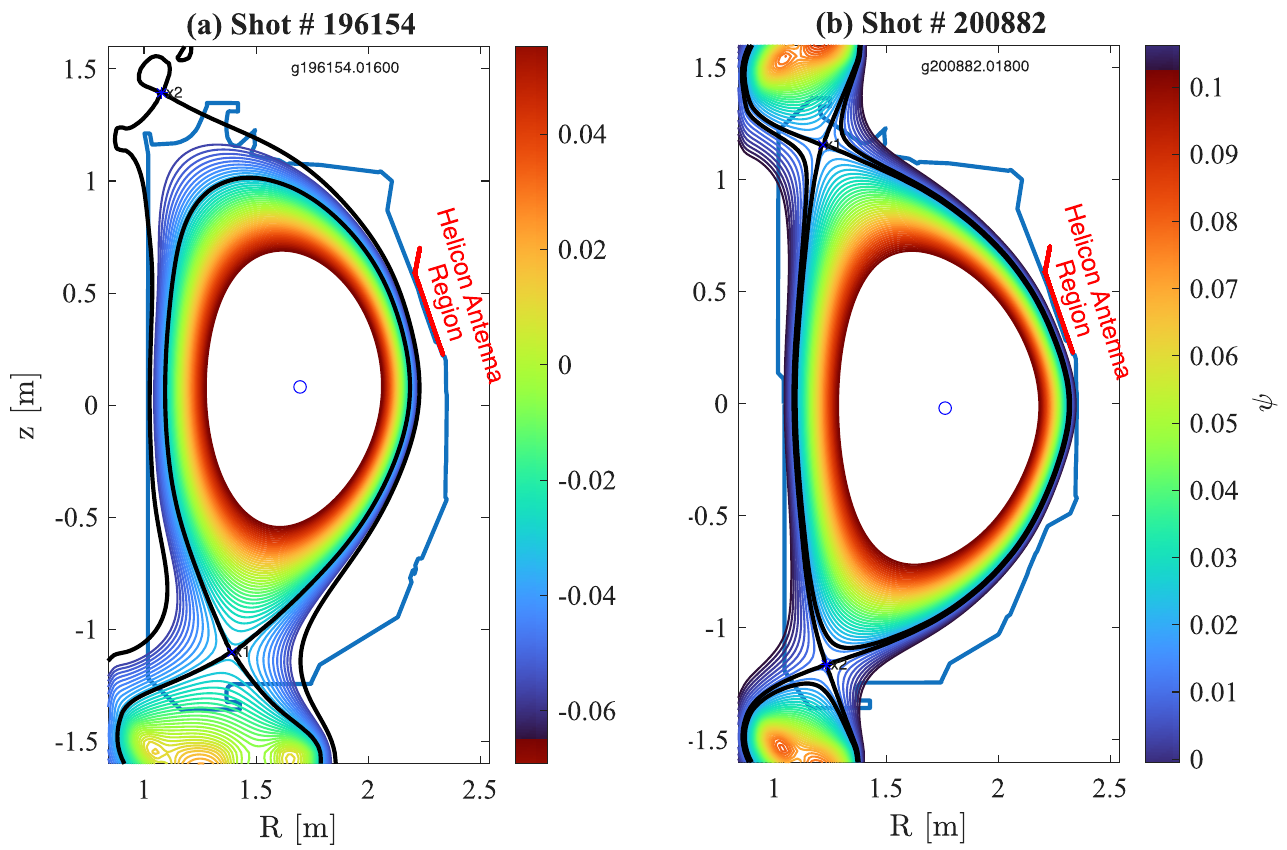}
    \caption{EFIT reconstructions of poloidal flux surfaces for two DIII-D discharges with different antenna-plasma gaps: \#196154 (larger gap $ \sim 7$~cm) and \#200882 (smaller gap $ \sim 4$~cm). The change in LCFS position relative to the helicon antenna modifies the sheath-connected surface area and plasma exposure, affecting RF sheath formation and impurity sourcing.}
    \label{fig:efitgap}
\end{figure}

EFIT reconstructions of the poloidal magnetic flux (Figure~\ref{fig:efitgap}a-b) demonstrate that discharge \#196154 featured a relatively larger antenna–plasma gap of approximately 7~cm, while in discharge \#200882, the last closed flux surface (LCFS) extends much closer to the antenna structure, reducing the gap to approximately 4~cm. This geometric variation directly influences the sheath-connected area, thereby modifying both the local plasma exposure and the likelihood of RF sheath formation. The helicon antenna region, including the walls surrounding the helicon antenna (WSHA), is indicated by the red contours in Figures~\ref{fig:efitgap}a-b. A noticeable geometric tilt is present at the upper portion of the WSHA, which does not align with the EFIT reconstruction and is therefore excluded from the impurity transport simulations discussed later in Section~\ref{sec:5} and shown in Figure~\ref{fig:geometry}b.

\begin{figure}
    \centering
    \includegraphics[width=\linewidth]{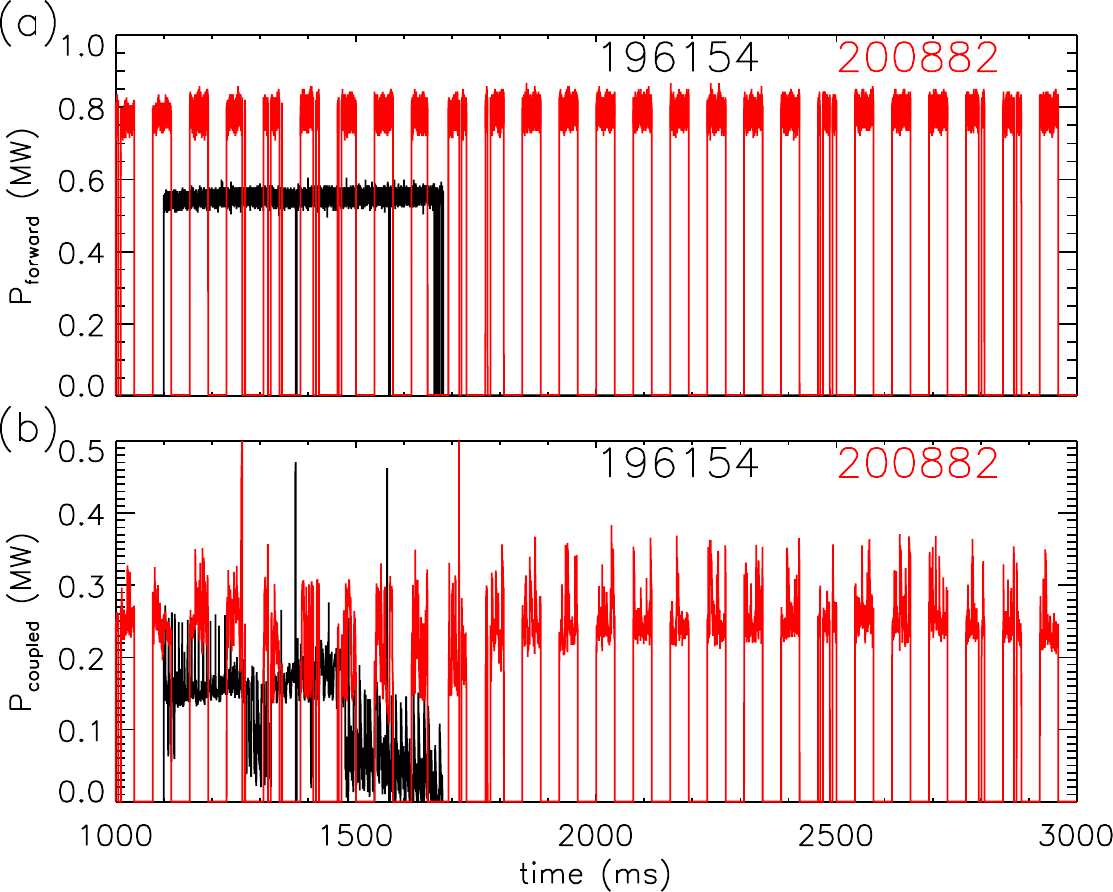}
    \caption{Time traces of (a) klystron forward power, $\mathrm P_{forward}$ (\texttt{HK1KDCFWD}) and (b) coupled power, $\mathrm P_{coupled}$ (\texttt{TWAPWRC}) for DIII-D discharges \#196154 (black) and \#200882 (red). Discharge \#196154 operated with higher forward power (550~kW) and a larger antenna–plasma gap, while \#200882 exhibited closer plasma contact but lower  power (800~kW). }
    \label{fig:rfpower}
\end{figure}

In addition to magnetic geometry, the discharges also differed in RF power coupling. Figure~\ref{fig:rfpower} shows time traces of klystron output and coupled power for both shots, based on the experimental signals \texttt{HK1KDCFWD} (klystron output helicon power, $\mathrm P_{forward}$) and \texttt{TWAPWRC} (coupled power, $\mathrm P_{coupled}$). Discharge \#200882 was driven at $\mathrm P_{forward} \sim 800$~kW and $\mathrm P_{coupled} \sim 250$~kW into the plasma, while the discharge \#196154 operated at $\mathrm P_{forward} \sim 550$~kW and $\mathrm P_{coupled} \sim 150$~kW. While both discharges showed moderate coupling compared to recent high-performance L-mode helicon experiments~\cite{Pinsker:2024}, the higher absolute coupled power in \#200882—together with its smaller antenna–plasma gap—provides a valuable case for evaluating sheath formation and impurity sourcing under distinct edge configurations.

\begin{figure*}
    \centering
    \includegraphics[width=\textwidth]{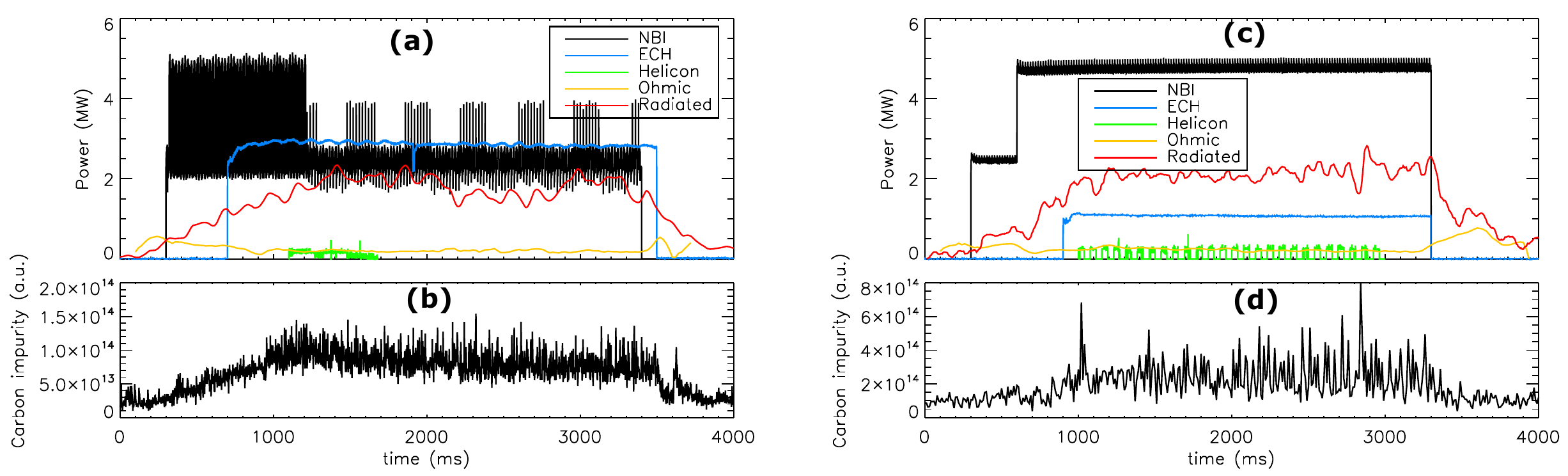}
    \caption{Time traces of power and carbon impurity signals. (a, c) Total injected power (NBI, ECH, Ohmic, and Helicon) and radiated power for discharges \#196154 and \#200882, respectively. Green trace shows helicon coupled RF power. (b, d) Corresponding time evolution of line-integrated carbon impurity emission. No significant rise in carbon levels is observed during helicon operation, indicating minimal global impurity response.}
    \label{fig:carbonprofile}
\end{figure*}

To assess the global impurity response to helicon operations, Figure~\ref{fig:carbonprofile} shows time traces of injected power components (NBI, ECH, Ohmic, Helicon, and total radiated) and the line-integrated carbon impurity content for both discharges. Subplots (a) and (c) correspond to the power balance for discharges \#196154 and \#200882, respectively, while (b) and (d) show the corresponding evolution of carbon impurity emission.

Despite the activation of the helicon antenna (green traces in (a) and (c)), the bottom panels (b, d) reveal \textit{no significant change in the global carbon impurity content} during helicon operation. For discharge \#196154 (b), the carbon emission remains steady throughout the RF pulse window, with only minor fluctuations. A similar trend is observed in \#200882 (d), where the impurity levels remain relatively unchanged even during the periods of elevated coupled helicon power.

These observations suggest that under the operating conditions in these H-mode discharges, \textit{the net carbon source driven by RF sheath effects is small compared to the background carbon level}, consistent with the minimal net erosion observed in simulations of \#196154 (see Section~\ref{sec:5}). Moreover, they indicate that the spatially localized erosion at the helicon antenna does not significantly alter the global carbon inventory, at least on the time scales of these discharges.

\subsection{SOLPS Background Plasma Profiles and Experimental Validations}

To provide physically realistic boundary conditions for sheath and impurity transport modeling, SOLPS-ITER (referred as simply SOLPS henceforth) background plasma profiles are generated for two DIII-D discharges: \#196154 and \#200882. These discharges were selected for their contrasting antenna–plasma gaps--approximately 7~cm and 4~cm, respectively--enabling a comparative evaluation of geometric and RF power coupling effects on local edge plasma conditions.

\begin{figure}
    \centering
    \includegraphics[width=\linewidth]{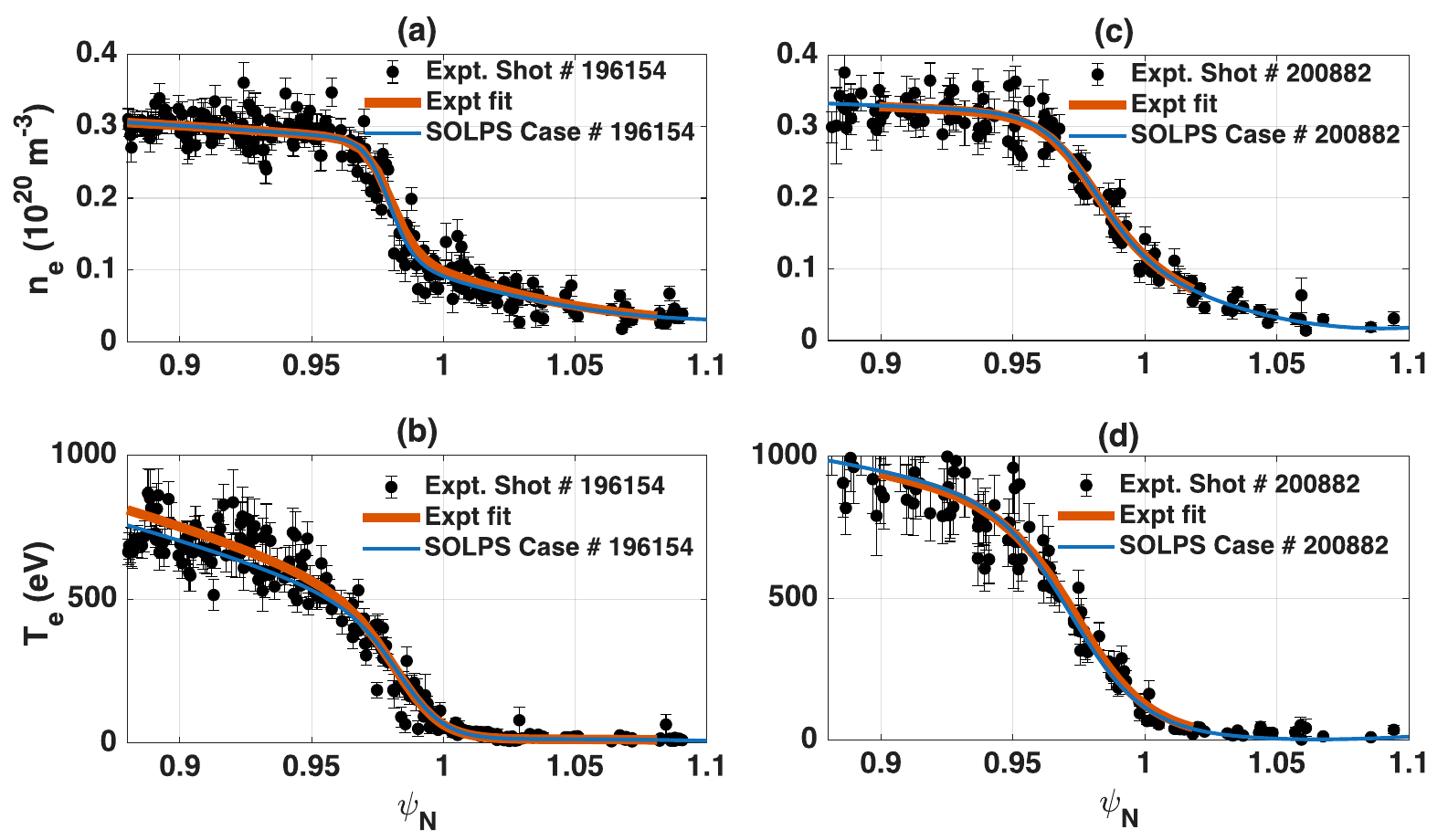}
    \caption{Comparison of SOLPS-ITER simulations and Thomson scattering measurements for DIII-D discharges \#196154 and \#200882. (a) Electron density $ n_e$ and (b) electron temperature $ T_e$ for discharge~\#196154; (c) $ n_e$ and (d) $ T_e$ for discharge~\#200882. Experimental data (black circles with error bars) are $ \tanh$ fit (orange), while SOLPS-ITER results are shown as blue curves. The profiles show strong consistency across the separatrix and into the SOL in both discharges.}
    \label{fig:profiles_combined}
\end{figure}

Figure~\ref{fig:profiles_combined} compares the SOLPS predictions of electron density ($ n_e$) and electron temperature ($ T_e$) to experimental measurements from Thomson scattering for both discharges. Subfigures~\ref{fig:profiles_combined}(a) and (b) show the $ n_e$ and $ T_e$ profiles, respectively, for discharge~\#196154, while subfigures~\ref{fig:profiles_combined}(c) and (d) show the same quantities for discharge~\#200882. Experimental data are shown as solid circles with error bars, and SOLPS solutions are plotted as smooth blue curves. In both cases, the SOLPS profiles were generated in interpretive mode by adjusting the cross-field transport coefficients ($ D_\perp$, $ \chi^{e,i}_\perp$) to match upstream experimental data using the iterative procedure described in references \citen{canik_2011} and \citen{Wilcox_NF_2025}. The resulting profiles exhibit very good agreement with experiment across the separatrix and into the SOL region.

Each simulation includes deuterium (D$^+$) and carbon ion species, with both physical sputtering and a fixed 2\% chemical sputtering yield applied along the divertor and outer SOL boundary.  For discharge~\#196154, the power balance used in SOLPS is $ P_{\mathrm{SOL}} = P_{\mathrm{NBI}} + P_{\mathrm{ECH}} + P_{\mathrm{Ohmic}} + P_{\mathrm{Helicon}} - P_{\mathrm{rad}}^{\mathrm{core}} = 7.6$~MW, with $ P_{\mathrm{Helicon}} = 0.15$~MW. For discharge~\#200882, $ P_{\mathrm{SOL}} = 7.1$~MW, with $ P_{\mathrm{Helicon}} = 0.25$~MW.

\begin{figure}
    \centering
    \includegraphics[width=\linewidth]{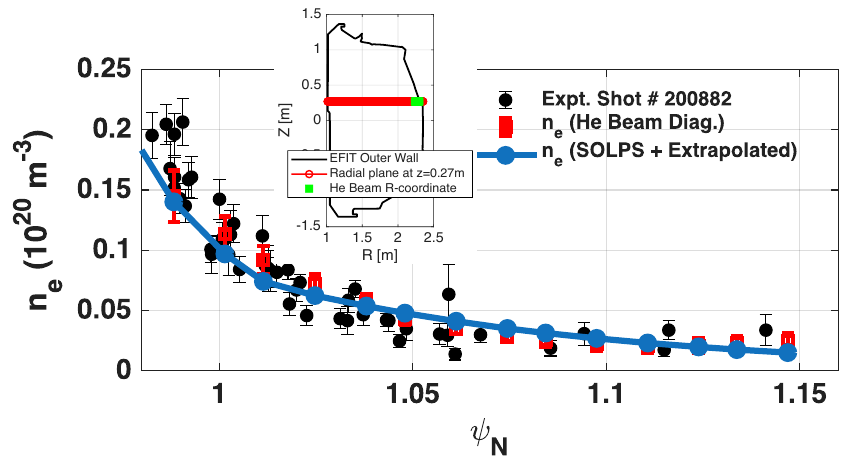}
    \caption{Comparison of electron density \( n_e\) in front of the helicon antenna for DIII-D discharge \#200882 from SOLPS (black curve with yellow circles) and He-beam diagnostic measurements (red boxes with error bars). The He-beam line-of-sight passes radially across the antenna aperture, providing a critical constraint on the local plasma conditions relevant to RF sheath formation and impurity erosion. Inset: geometry of the diagnostic view relative to the antenna face.}
    \label{fig:hebeam200882}
\end{figure}

To further constrain the  plasma profiles near the helicon antenna and in the WSHA region for discharge~\#200882, high-resolution  $ n_e$-data are obtained using the helium beam diagnostic (see \ref{app:heData} for details on the diagnostics), which provides 18 radial points across the antenna aperture at $ z = 0.27$~m with 3.7~mm resolution. Figure~\ref{fig:hebeam200882} compares the SOLPS + extrapolated $ n_e$ profile with both He-beam and Thomson data, showing good agreement across the entire domain. The inset illustrates the diagnostic's radial line-of-sight across the antenna face.

In regions beyond the SOLPS mesh--particularly near the outer wall, helicon antenna and the WSHA region--profiles were extrapolated using polynomial fits to ensure smooth and continuous boundaries. These validated and extended plasma backgrounds are used as inputs to the STRIPE framework for RF sheath potential calculations, sputtering yield evaluations, and whole-device impurity transport simulations.

\subsection{Carbon Charge State Densities from SOLPS}

\begin{figure}
    \centering
    \includegraphics[width=\linewidth]{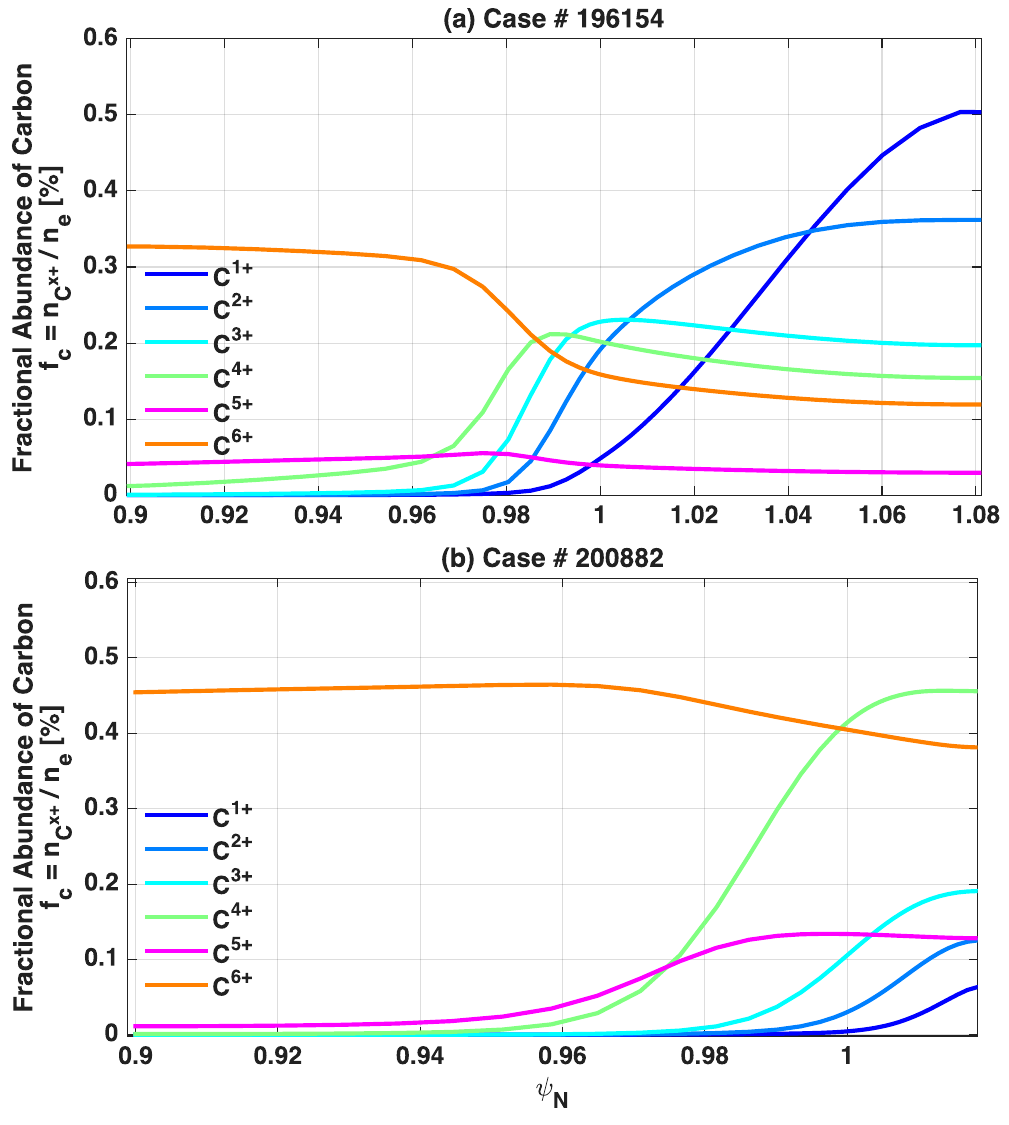}
    \caption{Percentage fractional abundances \( f_c = n_{C^{x+}} / n_e\) of carbon charge states computed by SOLPS-ITER for DIII-D discharges: (a) \#196154 and (b) \#200882. Each curve shows the contribution of a given charge state (\( C^{x+}:\,C^{1+}\)-\( C^{6+}\)) relative to the local electron density. Discharge \#200882 exhibits a strong peak in C\(^{6+}\), while \#196154 displays a broader distribution dominated by lower charge states.}
    \label{fig:carbon_fc}
\end{figure}

Figure~\ref{fig:carbon_fc} shows the fractional abundances of carbon charge states (\( C^{x+}\), where \( x = 1\)--6) predicted by SOLPS-ITER for DIII-D discharges \#196154 and \#200882. These are expressed as percentages relative to the local electron density: \( f_c = (n_{C^{x+}} / n_e) \times 100\). Since the structured SOLPS-ITER grid does not extend radially to PFCs, local erosion at material surfaces is not modeled explicitly. Instead, both chemical and physical sputtering sources are imposed along the outer radial ("North") boundary, with source rates based on local ion fluxes, plasma conditions, and surface temperatures. Chemical sputtering dominates under the cooler SOL conditions typical of DIII-D. The carbon impurities are then evolved using SOLPS's multi-fluid impurity model, which solves coupled continuity and momentum equations for each charge state, assuming ion temperatures equal to those of the main ions. Ionization and recombination rates are applied using ADAS data, allowing self-consistent prediction of charge state distributions across the edge.

In discharge~\#196154, a more gradual carbon ionization profile is observed. Lower charge states (\( C^{1+}\), \( C^{2+}\)) remain prevalent beyond the separatrix, while \( C^{6+}\) is significantly suppressed--consistent with the lower edge $ T_e$ shown in Figure~\ref{fig:profiles_combined}(a--b). Under these cooler conditions, carbon impurities are more likely to undergo recombination and charge exchange, and follow shorter transport paths due to reduced ionization lifetimes and frequent interactions with background neutrals.

In contrast, discharge~\#200882 exhibits deeper ionization, with \( C^{6+}\) dominating across a wide range of normalized flux \( \psi_N\), extending well into the SOL. This reflects elevated edge \( T_e\) (Figure~\ref{fig:profiles_combined}(c--d)) and more complete ionization in the edge. The presence of fully stripped carbon ions near the helicon antenna is particularly relevant, as higher charge states experience stronger acceleration in RF sheaths due to their larger charge-to-mass ratio. 

The predicted charge state balance is also sensitive to assumptions about the main ion parallel flow, which influences impurity retention and re-ionization pathways. A more detailed discussion of this sensitivity and its implications for modeling in helicon geometries is provided in section~\ref{sec:discussion2}.

Overall, this comparison illustrates how edge plasma conditions directly shape the local carbon ionization balance. Because charge state affects impurity transport, radiation losses, and sheath interactions, accurate modeling of carbon charge state evolution is always important for predictive PMI assessments in helicon-driven scenarios.

\section{Additional Inputs to the STRIPE Framework}
\label{sec:4}
\subsection{Geometry Defeaturing}
\label{subsec:41}
\begin{figure}
    \centering
    \includegraphics[width=\linewidth]{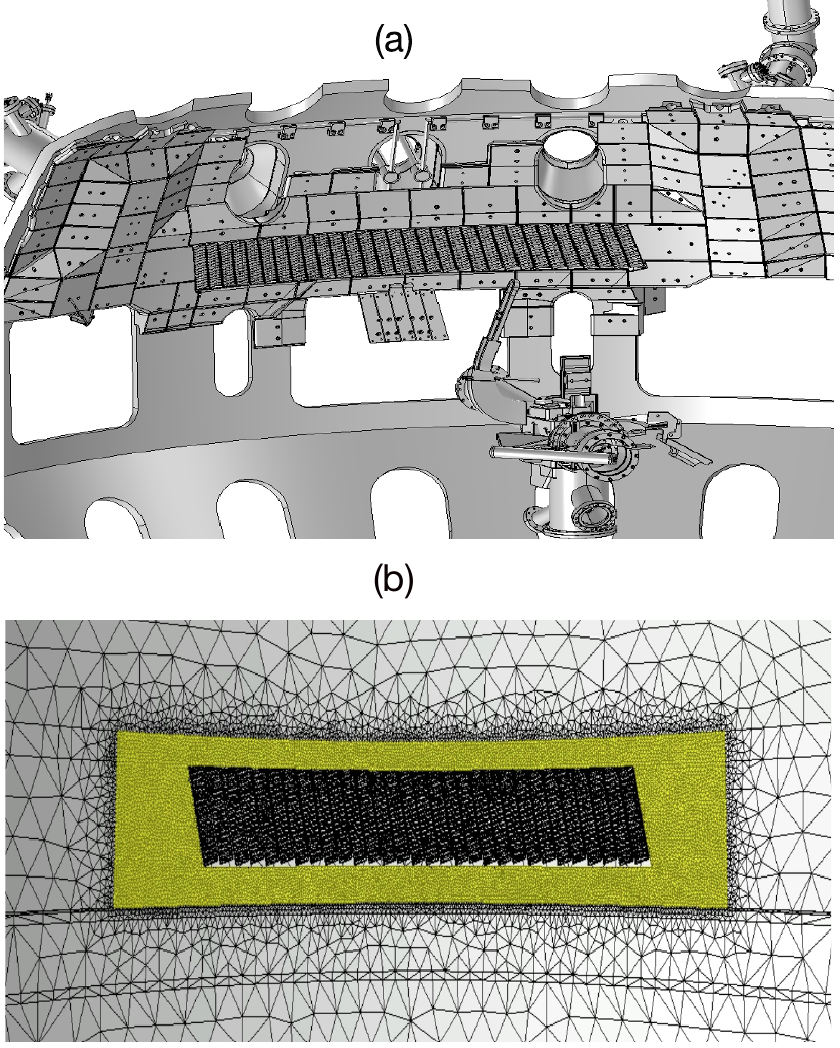}
    \caption{(a) Engineering CAD model of the DIII-D helicon antenna, illustrating the detailed mechanical design and integration with surrounding structures. (b) The defeatured geometry prepared for finite-element simulations in COMSOL and impurity transport modeling with GITRm retains essential antenna surfaces and plasma-facing components while removing small-scale features to reduce computational complexity.}
    \label{fig:geometry}
\end{figure}

The DIII-D helicon antenna is a 30-element comb-line traveling-wave structure with a toroidal width of 1.5\,m, designed to couple up to 1\,MW of RF power at 476\,MHz with high load resilience~\cite{Compernoll:2021, Pinsker:2024}. In preparing this geometry for simulation, we started from an extremely complex engineering CAD model comprising over 600 individual parts, as illustrated in Fig.~\ref{fig:geometry}(a). This detailed representation included numerous mechanical features such as nuts, bolts, fastener and cooling channels etc. While some of these are essential for fabrication and assembly, they may not be important for computational modeling of RF sheath potentials and impurity transport. Defeaturing this geometry, shown in Fig.~\ref{fig:geometry}(b), was a necessary and time-intensive process to enable robust meshing and efficient finite-element calculations in COMSOL, as well as subsequent impurity transport simulations with GITR/GITRm. The simplified model retains all plasma-facing and RF-relevant surfaces, including the Faraday screen and current straps, while removing small-scale details that do not significantly impact sheath behavior but would otherwise impose prohibitive computational overhead.

\subsection{RF Sheath Simulations with COMSOL}
\label{subsec:42}

The influence of RF sheath effects on wave propagation and rectified DC potential formation at the DIII-D's WSHA region during the helicon operation is modeled using the sheath-equivalent dielectric layer technique~\cite{Beers1,Beers2,Tierens_2024}. In this approach, thin dielectric layers introduced on WSHA, as shown by green region  in Figure~\ref{fig:geometry}b,  mimic the behavior of RF sheaths by incorporating spatially varying permittivity $\varepsilon$ and conductivity $\sigma$. These properties are derived from a sheath impedance model~\cite{myra2017physics}, which depends on local plasma parameters including $ n_e$, $ T_e$, magnetic field orientation, and the local RF potential across the sheath layer. The effective permittivity $\varepsilon_{\mathrm{eff}}$ and conductivity $\sigma_{\mathrm{eff}}$ are computed from the sheath impedance $Z_{\mathrm{sh}}$ and implemented in COMSOL via a custom material definition. The resulting rectified DC sheath potential $ V_{\mathrm{sheath}}$ is then obtained self-consistently from the wave solution.

The sheath model approximates the physical sheath as a thin artificial dielectric layer of thickness $d = 2$~mm, satisfying $\lambda_{D} \ll d \ll \lambda_{\mathrm{RF}}$, where $\lambda_{D}$ is the Debye length and $\lambda_{\mathrm{RF}}$ is the RF wavelength. This formulation embeds sheath physics directly into the material properties, thereby avoiding the need for explicit sheath boundary conditions as required in other sheath models such as Petra-M~\cite{shiraiwa2023magnetic} or Stix~\cite{migliore2024development}.

A four-step iterative procedure is employed:
\begin{enumerate}
    \item Run an initial COMSOL simulation assuming vacuum sheaths ($\sigma = 0$, $\varepsilon = 1$).
    \item Compute updated $Z_{\mathrm{sh}}$ and the corresponding $\varepsilon_{\mathrm{eff}}$, $\sigma_{\mathrm{eff}}$.
    \item Update the material properties in COMSOL.
    \item Repeat until $Z_{\mathrm{sh}}$ converges (typically within 3-5 iterations).
\end{enumerate}

\begin{figure}
    \centering
    \includegraphics[width=\linewidth]{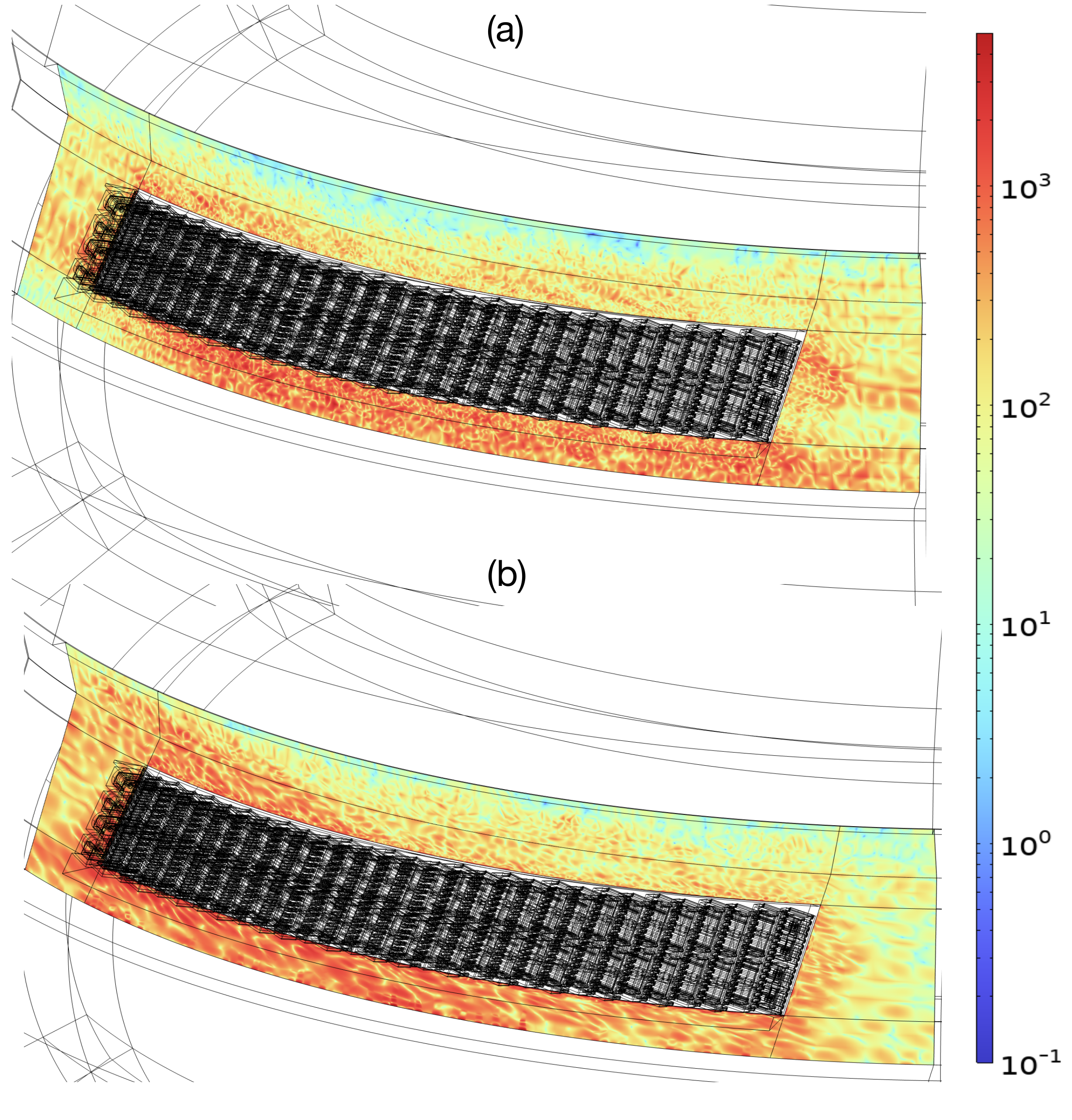}
    \caption{Computed rectified DC sheath potentials ($V_{\mathrm{sheath}}$) on surfaces surrounding the DIII-D helicon antenna from COMSOL simulations. (a) Discharge \#196154, with a larger antenna–plasma gap, shows sheath potentials primarily concentrated near the bottom portion of the WSHA. (b) Discharge \#200882, with a reduced gap, exhibits stronger and more broadly distributed potentials across the entire WSHA. The spatial structure of $V_{\mathrm{sheath}}$ reflects the antenna geometry, with poloidal modulation imposed by the Faraday screen and toroidal variations associated with the antenna modules.}
    \label{fig:VDC}
\end{figure}

 Figure~\ref{fig:VDC} presents the computed rectified DC  $V_{\mathrm{sheath}}$ for two representative discharges. In discharge~\#196154, the sheath potentials are concentrated near the bottom portion of the antenna, while in discharge~\#200882, they are more broadly distributed across the WSHA region. In both cases, the poloidal structure of $V_{\mathrm{sheath}}$ reflects modulation by the Faraday screen, while toroidal variations are associated with individual antenna modules. The localized enhancement of $V_{\mathrm{sheath}}$ in the lower region of both discharges is attributed to reduced local $ n_e$, which enhances sheath rectification in regions farther from the plasma edge. These sheath-aware boundary conditions are subsequently passed to the STRIPE framework to compute RF-enhanced sputtering and impurity transport, as described in Section~\ref{subsec:41}.

\begin{figure}
    \centering
    \includegraphics[width=0.7\linewidth]{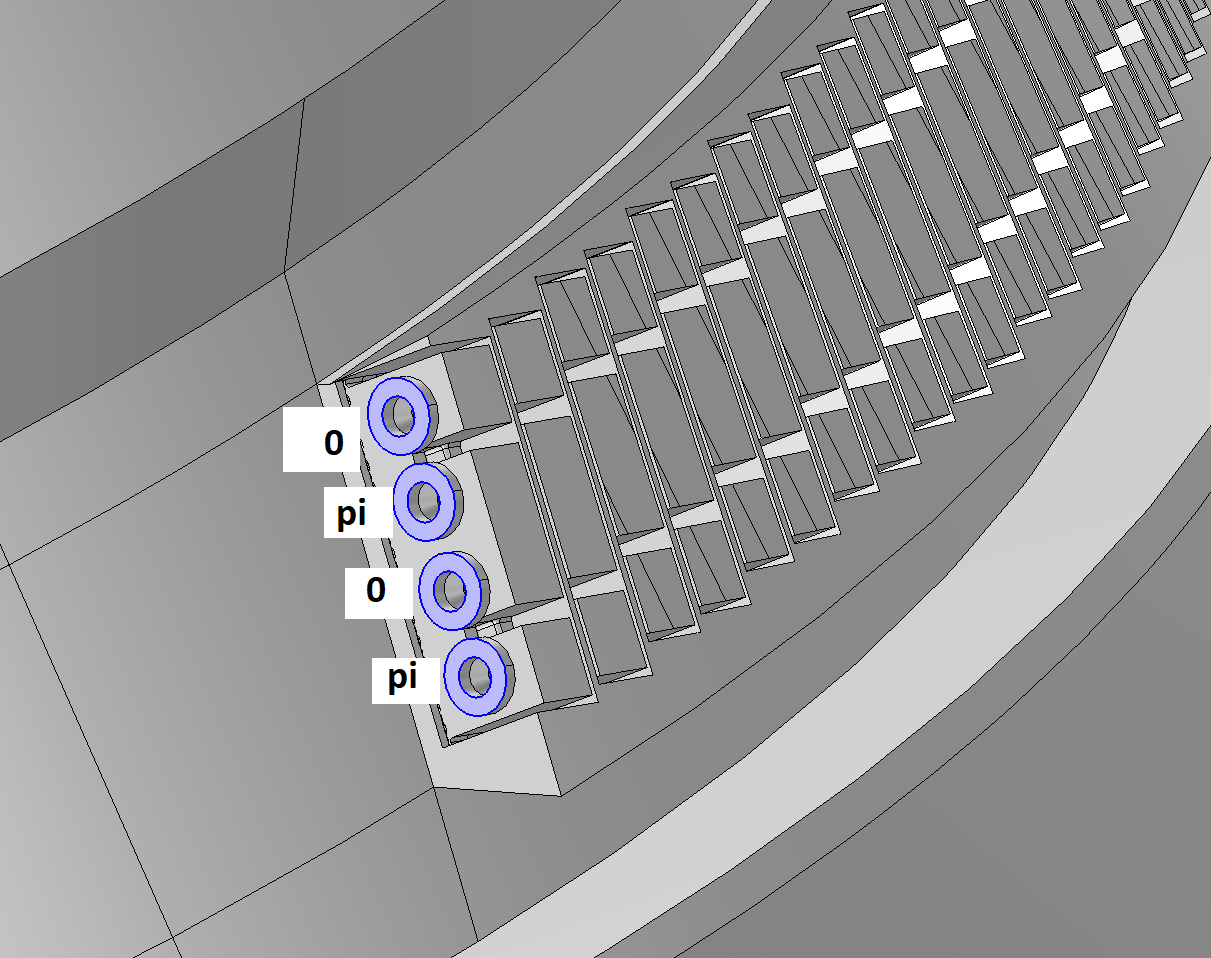}
    \caption{The four powered ports (of eight ports total), with phasing $0-\pi-0-\pi$.}
    \label{fig:phasing}
\end{figure}

Unlike in a related work \cite{Tierens_2024} where core absorption is handled using Perfectly Matched Layers, here absorption is entirely due to a collisionality added to the cold plasma. For the reference case, 360kW of power is coupled to the collisional absorption in the plasma. Of the 8 ports in the model, 4 are powered with phasing $0-\pi-0-\pi$ \cite{Compernoll:2021} as shown in Figure \ref{fig:phasing}.

\begin{figure}
    \centering
    \includegraphics[width=1.0\linewidth]{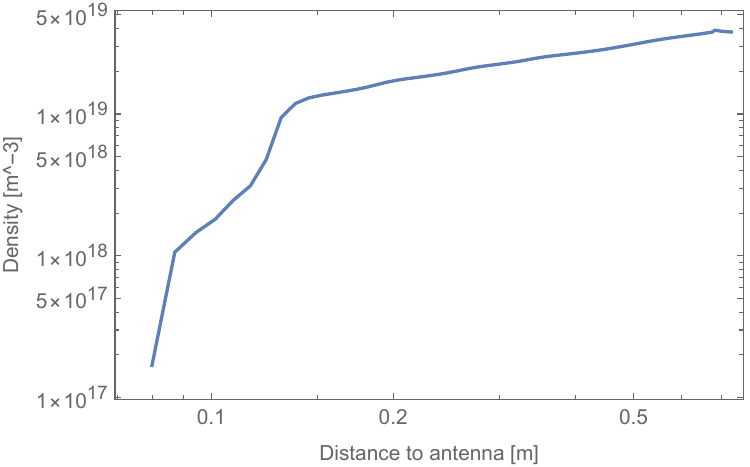}
    \includegraphics[width=1.0\linewidth]{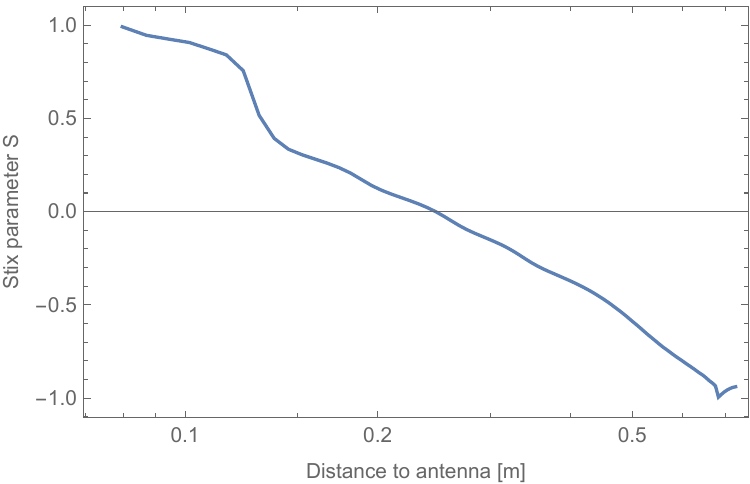}
    \caption{Top: reference density profile, as a function of the distance to the antenna aperture. Bottom: corresponding Stix parameter S.}
    \label{fig:diiidn}
\end{figure}

The reference density profile is shown in Figure \ref{fig:diiidn}. With this density profile, a magnetic field strength of $B=1.5$T (assumed constant, no $1/R$ variation), and a wave frequency of 476MHz, the Lower Hybrid Resonance $S=0$ is in front of the antenna.

\begin{figure}
    \centering
    \includegraphics[width=1.0\linewidth]{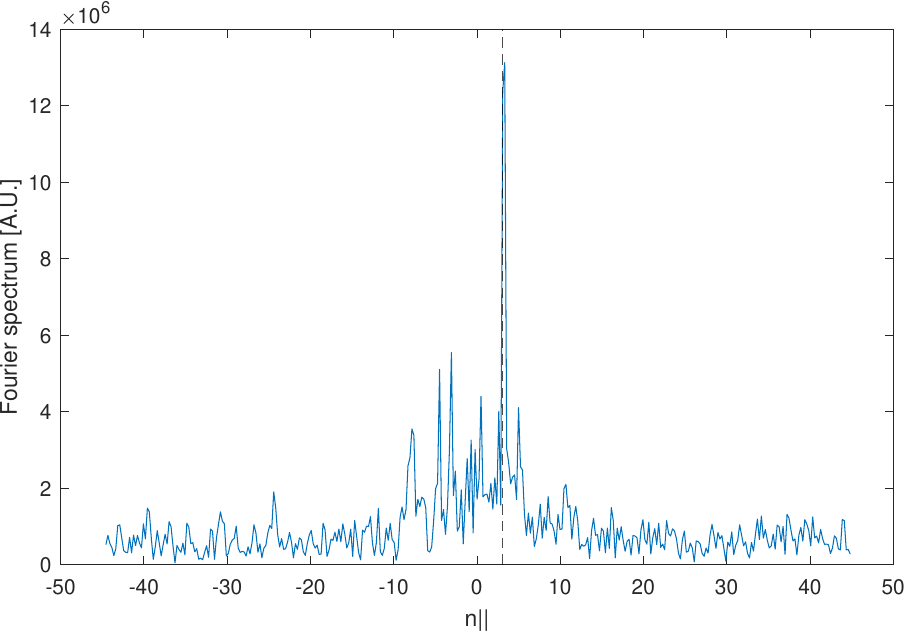}
    \caption{Numerical Fourier spectrum of the vertical component of the RF electric field. The peak is at $n_{\parallel}=3$, which matches the expected value\cite{lau2021helicon}. Note also that the spectrum is asymmetric, consistent with it being a traveling wave antenna.}
    \label{fig:numericalSpectrum}
\end{figure}

Earlier 2D modeling efforts~\cite{lau2021helicon} typically imposed a fixed parallel wavenumber, often using \(n_{\parallel} = 3\), to characterize the excited RF spectrum. In our 3D simulations, however, \(n_{\parallel}\) is determined self-consistently from the antenna geometry and plasma environment. As shown in Figure~\ref{fig:numericalSpectrum}, the numerical field spectrum indeed peaks near \(n_{\parallel} = 3\), consistent with the nominal helicon coupling condition.

\begin{figure}
    \centering
    \includegraphics[width=1.0\linewidth]{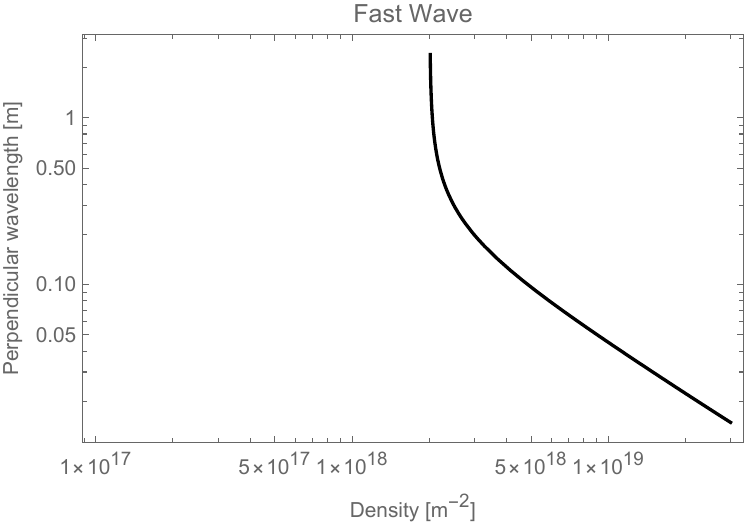}
    \includegraphics[width=1.0\linewidth]{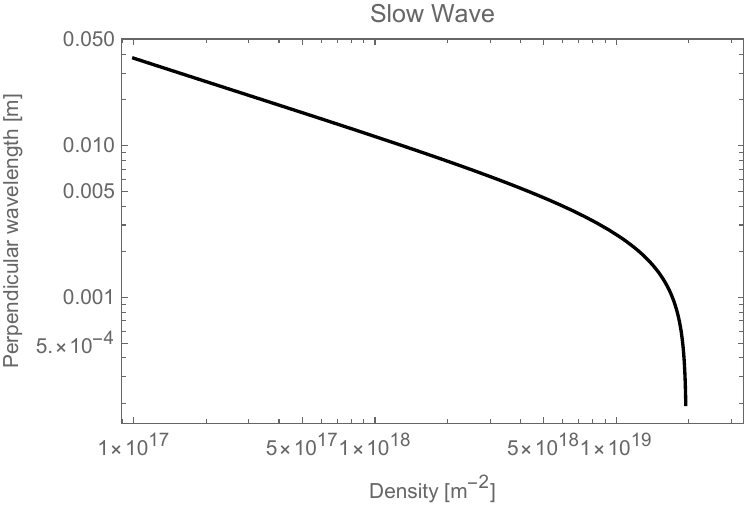}
    \caption{Perpendicular (radial) wavelengths of the Fast and Slow waves assuming $n_{\parallel}=3$ and $n_{\mathrm{poloidal}}=0$ The slow wave propagates in the low density edge plasma, and the fast wave must cross an evanescent layer before becoming propagative.}
    \label{fig:dispersion}
\end{figure}

The cold plasma dispersion relations for the fast and slow wave branches are presented in Figure~\ref{fig:dispersion}. As expected, the fast wave crosses an evanescent region before becoming propagative. In contrast, the slow wave is propagative over a substantial edge layer (approximately 10~cm thick) in front of the antenna and becomes evanescent at higher densities. To compute the slow wave dispersion, we adopt \(n_{\parallel} = 3\), consistent with the fast wave and common practice~\cite{lau2021helicon}. However, the actual slow wave spectrum is expected to include higher \(n_{\parallel}\) components and shorter perpendicular wavelengths, particularly near material surfaces, as shown in recent work~\cite{tierens2024slow}. The assumption of a vanishing vertical wavenumber (\(k_z = 0\)) may also not hold for slow wave components. These important spectral and modeling considerations are discussed further in section~\ref{sec:6.2}.

\subsection{Sputtering Yield Calculations with RustBCA }
\label{subsec:43}

$ Y (\mathcal{E, \theta})$ (simply referred as $ Y$ henceforth) is pre-computed with RustBCA for monoenergetic, angle-resolved impacts of neutral $ D$ and $ C$ projectiles on $ C$ targets. Simulations spanned $ \mathcal{E}$ from 70~eV to 10~keV and $ \theta$ from 0° (normal) to 90° (grazing). Like other BCA codes, RustBCA assumes neutral projectiles and neglects long-range Coulomb interactions or charge-state-dependent stopping, an approximation generally reasonable near material surfaces where ions neutralize prior to impact.

\begin{figure}
    \centering
    \includegraphics[width=\linewidth]{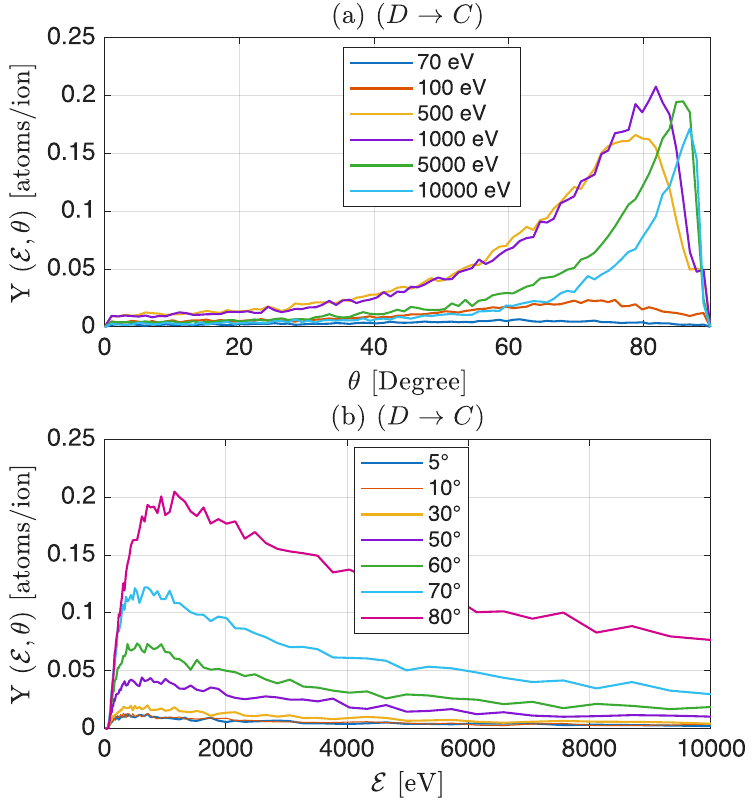}
    \caption{$ Y(\mathcal{E}, \theta)$ for $ D$ on $ C$ computed with RustBCA. 
    (a) $ Y$ vs.\ $ \theta$ for selected $ \mathcal{E}$: 70~eV (red), 100~eV (orange), 200~eV (gold), 500~eV (purple), 1~keV (green), 5~keV (blue), 10~keV (cyan). 
    (b) $ Y$ vs.\ $ \mathcal{E}$ for selected $ \theta$: 0°, 10°, 30°, 50°, 60°, 70°, 80°.}
    \label{fig:rustbca_d_on_c}
\end{figure}

\begin{figure}
    \centering
    \includegraphics[width=\linewidth]{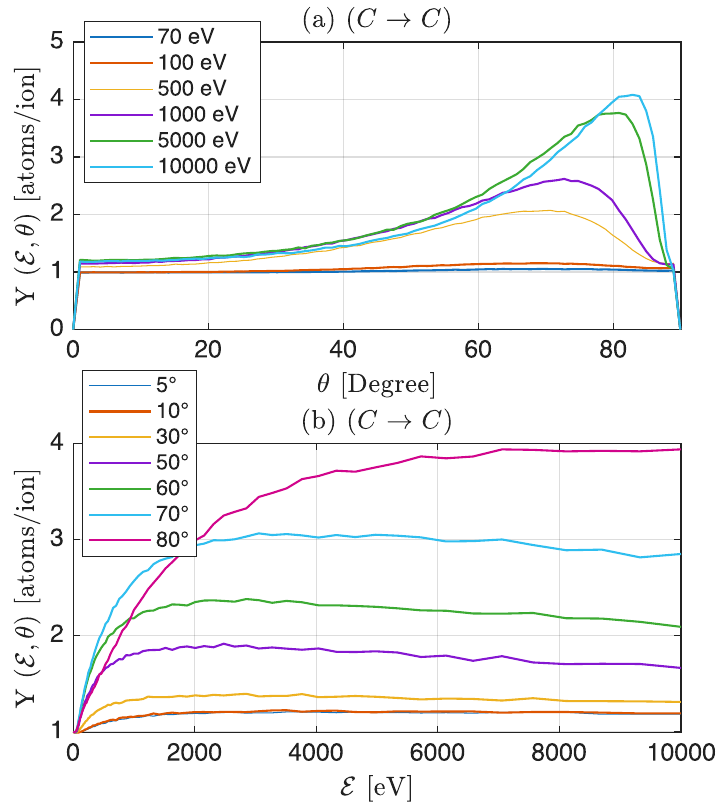}
    \caption{$ Y(\mathcal{E}, \theta)$ for $ C$ on $ C$ computed with RustBCA. 
    (a) $ Y$ vs.\ $ \theta$ for selected $ \mathcal{E}$: 70~eV (red), 100~eV (orange), 200~eV (gold), 500~eV (purple), 1~keV (green), 5~keV (blue), 10~keV (cyan). 
    (b) $ Y$ vs.\ $ \mathcal{E}$ for selected $ \theta$: 0°, 10°, 30°, 50°, 60°, 70°, 80°.}
    \label{fig:rustbca_c_on_c}
\end{figure}

Figures~\ref{fig:rustbca_d_on_c} (a-b) and~\ref{fig:rustbca_c_on_c}(a-b) show the dependence of $ Y$ on $ \theta$ and $ \mathcal{E}$ for $ D$ and $ C$ projectiles, respectively. For $ D$ on $ C$, $ Y$ increases gradually with $ \mathcal{E}$ up to ~1-1.5~keV, then decreases across all $ \theta$ at higher $ \mathcal{E}$. This non-monotonic trend likely results from deeper penetration of light $ D$ ions at multi-keV energies, reducing near-surface energy deposition. Above 2~keV, $ Y$ remains <0.2 atoms/ion for all $ \theta$. Notably, above $\sim$2~keV, yields remain low at near-normal incidence and increase strongly with oblique incidence, peaking at large $\theta$ ($\gtrsim 80^\circ$)..
d
In contrast, $ C$ self-sputtering shows a strong, largely monotonic dependence. $ Y$ increases steeply with $ \mathcal{E}$ up to ~2~keV, then tends to saturate for most $ \theta$. Beyond 2~keV, further increases produce only modest additional erosion except at $ \theta \approx 70$-80°, where $ Y$ continues rising to ~3 atoms/ion at 10~keV. For all $ \mathcal{E}$, $ Y$ is lowest at normal $ \theta$ and increases with obliquity, peaking before decreasing near true grazing impact. This reflects the balance between near-surface cascades and reflection losses, with maximum $ Y$ at oblique $ \theta$ and moderate to high $ \mathcal{E}$.

\begin{figure}
    \centering
    \includegraphics[width=\linewidth]{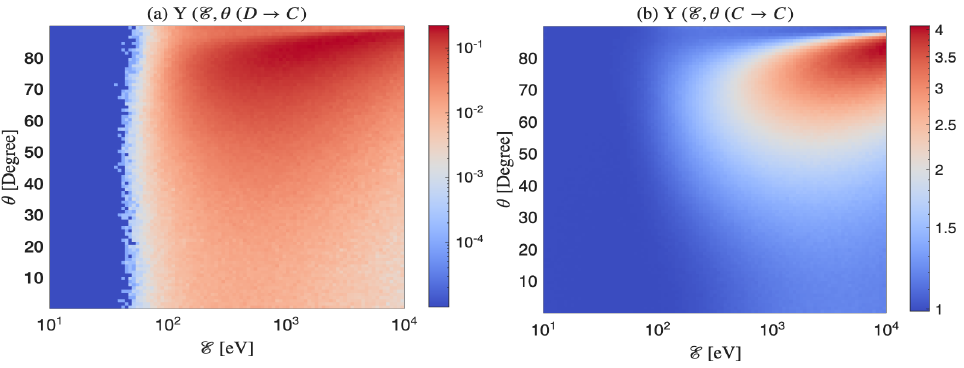}
    \caption{Two-dimensional $ Y(\mathcal{E}, \theta)$ surfaces computed with RustBCA. For $ C$ projectiles, $ Y$ increases strongly with $ \mathcal{E}$ up to ~2~keV and saturates thereafter, especially at oblique $ \theta$. For $ D$, $ Y$ peaks at intermediate $ \mathcal{E}$ and declines at higher energies.}
    \label{fig:rustbca_2d}
\end{figure}

Figure~\ref{fig:rustbca_2d} shows two-dimensional $ Y(\mathcal{E}, \theta)$ surfaces. Under helicon antenna conditions, where rectified RF sheath potentials can accelerate ions into the multi-keV range, these results indicate that $ D$-induced erosion remains modest and decreases at high $ \mathcal{E}$, while energetic $ C$ contributes disproportionately, especially at oblique $ \theta$. 

\section{Results: Local Carbon Erosion and Whole-Device Transport from the Helicon Antenna}
\label{sec:5}

This section presents simulation results of carbon erosion and impurity transport originating from the helicon antenna in two DIII-D discharges. Using the STRIPE framework, we quantify local erosion patterns, net re-deposition, and global impurity distribution across the device. The analysis highlights how differences in antenna–plasma gap and edge plasma conditions influence RF sheath-driven erosion and core impurity retention.

\subsection{Effective Sputtering Yields and Gross Erosion}

The effective sputtering yield (Y$_\mathrm{eff}$), as discussed earlier in Section~\ref{sec:2}, quantifies the average number of carbon atoms ejected per incident ion. It is computed by integrating energy- and angle-resolved yields from RustBCA, weighted by the IEADs obtained using the GITR code. These 3D IEADs, evaluated at the WSHA region, are shaped by rectified RF sheath potentials from COMSOL (see Figure~\ref{fig:VDC}) and by the spatial distribution of incident ion fluxes derived from 2D SOLPS simulations for both discharges, which are extrapolated onto the 3D WSHA geometry assuming toroidal symmetry. In both discharges, sheath potentials locally are in the range of  1-5~keV, accelerating ions to energies sufficient to enhance sputtering at PFCs. The resulting Y$_\mathrm{eff}$ inherently incorporates effects due to geometric curvature, local ion-sheath interactions, and the magnetic field incidence angle to the material surface.

\begin{figure*}
\centering
\includegraphics[width=\textwidth]{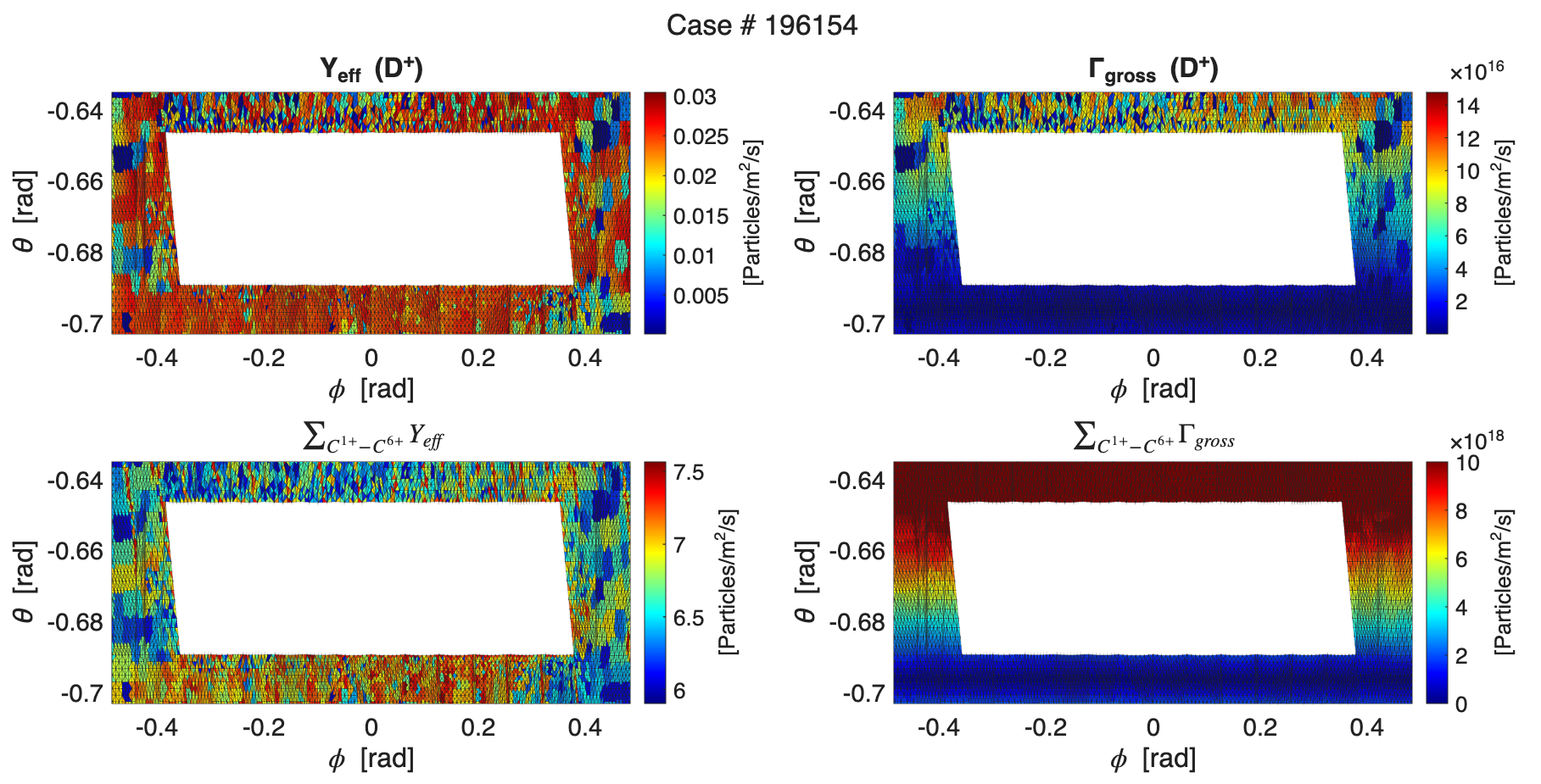}
\caption{Spatially resolved effective sputtering yields ($Y_\mathrm{eff}$) and gross carbon erosion fluxes ($\Gamma_\mathrm{gross}$) for DIII-D discharge \#196154. Top row: $Y_\mathrm{eff}$ and $\Gamma_\mathrm{gross}$  to D$^+$ ions. Bottom row: $Y_\mathrm{eff}$ and $\Gamma_\mathrm{gross}$ summed over C$^{1+}$–C$^{6+}$ ions. Elevated sputtering yields are observed near the bottom portion of the WSHA region due to strong RF sheath potentials. However, the gross erosion flux in this region remains weak due to reduced plasma flux, driven by a relatively  larger antenna–plasma gap in that region. Both D$^+$ and carbon ion-induced erosion are strongest in the upper to midplane regions ($-0.67 < \theta < -0.64$~rad), where high ion fluxes coincide. }
\label{fig:total_gross_196154}
\end{figure*}

\begin{figure*}
\centering
\includegraphics[width=\textwidth]{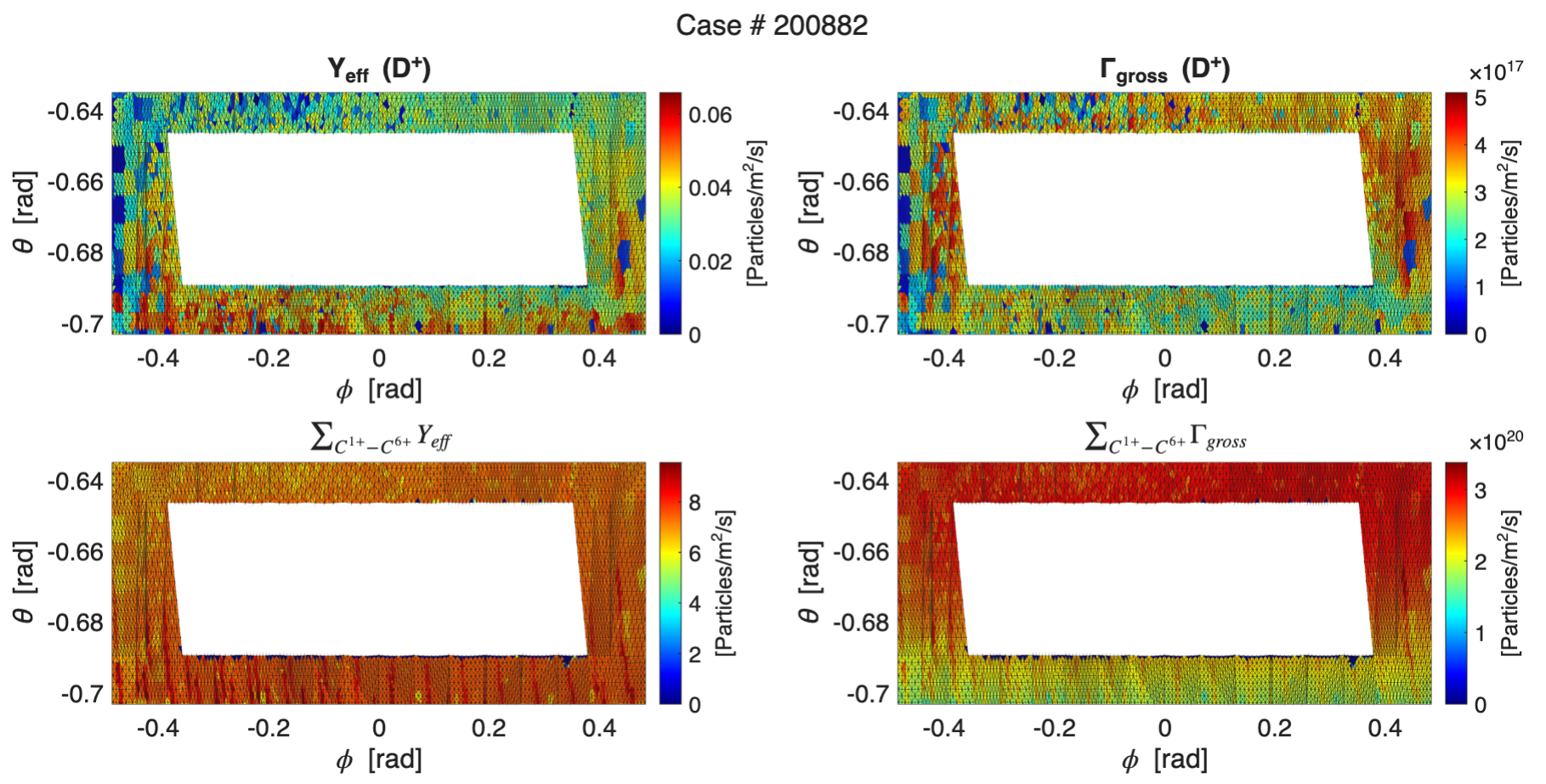}
\caption{Spatially resolved effective sputtering yields ($Y_\mathrm{eff}$) and gross carbon erosion fluxes ($\Gamma_\mathrm{gross}$) for DIII-D discharge \#200882. Top row: $Y_\mathrm{eff}$ and $\Gamma_\mathrm{gross}$ due to D$^+$ ions. Bottom row: $Y_\mathrm{eff}$ and $\Gamma_\mathrm{gross}$ summed over C$^{1+}$–C$^{6+}$ ions. Both D$^+$ and carbon ion-induced erosion are strongest in the upper to midplane regions ($-0.67 < \theta < -0.64$~rad), where high ion fluxes coincide. }
\label{fig:total_gross_200882}
\end{figure*}

Figures~\ref{fig:total_gross_196154} and \ref{fig:total_gross_200882} show the spatially resolved gross carbon erosion fluxes and Y$_\mathrm{eff}$  maps for both D$^+$ and carbon ions. The erosion flux  maps represent the product of sputtering yields and local plasma fluxes, summed over all six charge states of carbon, revealing how antenna geometry and edge plasma conditions shape material erosion patterns. For improved visualization, the 3D wall surface coordinates $(X, Y, Z)$ were transformed into cylindrical coordinates using $R = \sqrt{X^2 + Y^2}$, and angular variables were defined as $\phi = \tan^{-1}(Y/X)$ and $\theta = \tan^{-1}((Z - z_0)/(R - y_0))$, where $(y_0, z_0) = (2.21\mathrm m, 0)$ denotes the reference location for the transformation, corresponding to the approximate toroidal center of the helicon antenna. Here, $\phi$ represents the toroidal angle around the tokamak axis, while $\theta$ serves as an effective poloidal angle in the $R$–$Z$ plane along the antenna surface.

In discharge \#196154 (Figure~\ref{fig:total_gross_196154}), the antenna–plasma gap is approximately 7~cm, resulting in reduced plasma exposure to antenna-facing structures. The Y$_\mathrm{eff}$ for both D$^+$ and carbon ions is elevated near the bottom portion of the WSHA, consistent with stronger sheath potentials in that region predicted by COMSOL. However, the ion flux in this area is substantially lower, and the corresponding $\Gamma_\mathrm{gross}$ map remains weak. This is consistent with EFIT reconstructions (Figure~\ref{fig:efitgap}), which show the bottom section is farther from the LCFS, leading to reduced plasma density and ion access. Although lower density can enhance sheath rectification, as shown in Figure~\ref{fig:VDC}a, it does not result in significant erosion unless accompanied by sufficient ion flux. As a result, despite elevated sputtering yields at the bottom, peak erosion is observed in the upper to midplane regions of the WSHA ($-0.67 < \theta < -0.64$~rad), where plasma flux and sheath potentials are more favorably aligned.

Quantitatively, Y$_\mathrm{eff}$(D$^+$) remains low across the surface, with a maximum of approximately $\sim$0.03 atoms/ion, and the corresponding gross erosion flux, $\Gamma_\mathrm{gross}$(D$^+$), stays below $1.5 \times 10^{17}$~particles/m$^2$/s. In contrast, the effective sputtering yield for carbon ions is significantly higher--reaching up to $\sim$7.5 atoms/ion--primarily due to the dominance of carbon self-sputtering at sheath-accelerated energies in the keV range. The resulting total gross erosion flux for all the carbon charge states, $\sum \Gamma_\mathrm{gross}$(C$^{1+}$–C$^{6+}$), peaks at approximately $1 \times 10^{19}$~particles/m$^2$/s, nearly two orders of magnitude higher than the D$^+$-driven erosion flux. This contrast highlights the significant role of background impurities in driving material erosion, suggesting that a cleaner plasma environment with reduced impurity content would lead to substantially lower erosion rates.

In discharge \#200882 (Figure~\ref{fig:total_gross_200882}), the antenna–plasma gap is reduced to approximately 4~cm, increasing plasma exposure and resulting in more pronounced erosion. The $Y_\mathrm{eff}$ and $\Gamma_\mathrm{gross}$ maps for both D$^+$ and carbon ions exhibit similar spatial localization, with peak erosion again centered in the upper to midplane region of the antenna surface ($-0.67 < \theta < -0.64$~rad). This region experiences elevated ion flux due to its closer proximity to the LCFS, which enhances erosion activity. For D$^+$, Y$_\mathrm{eff}$ increases to approximately $\sim$0.06 atoms/ion, and the corresponding $\Gamma_\mathrm{gross}$(D$^+$), peaks at $5 \times 10^{17}$~particles/m$^2$/s—nearly three times higher than in discharge \#196154. Carbon ion-induced erosion also intensifies substantially. In this case, the effective sputtering yield for carbon reaches slightly above $\sim$8 atoms/ion, representing a moderate increase over \#196154. This moderate increase is attributed to sheath voltages remaining in similar keV range, where carbon self-sputtering approaches saturation beyond $\sim$2-4~keV depending on the incident angle with respect to the surface normal as can be inferred from Figure~\ref{fig:rustbca_c_on_c}b. 

Furthermore, the spatial distribution of Y$_\mathrm{eff}$(C) is broader, extending across a wider poloidal region of the antenna surface. This happens due to  the elevated sheath conditions being uniformly distributed as can be seen in Figure~\ref{fig:VDC}b, leading to enhanced carbon self-sputtering over a larger surface area. As a result, the total gross erosion flux from carbon, $\sum \Gamma_\mathrm{gross}$(C$^{1+}$–C$^{6+}$), reaches a peak of $3 \times 10^{20}$~particles/m$^2$/s--more than an order of magnitude higher than in the low-exposure case in the discharge \# 196154. These observations emphasize that while sputtering yields for carbon are similar or only modestly enhanced, it is the \textit{co-localization of strong ion flux and elevated sheath potentials}, combined with a \textit{broader spatial extent of energetic carbon impact}, that drives the significantly higher gross erosion in this configuration.

These comparisons highlight the nonlinear coupling between RF sheath enhancement, which governs Y$_\mathrm{eff}$, and edge plasma accessibility, which determines ion flux. High sputtering yields alone are insufficient to drive substantial erosion unless they coincide with strong plasma flux. Conversely, high local flux can only result in significant erosion if sheath potentials are sufficient to accelerate ions into the sputtering regime. 

\subsection{Net Erosion and Re-deposition Patterns from GITRm}
\label{subsec:43}

Net erosion and re-deposition of carbon impurities are modeled using the GITRm code (see \ref{app:gitrm}), which simulates full-orbit impurity transport in 3D geometries while accounting for sheath electric fields, magnetic geometry, charge state evolution, local re-deposition, and carbon self-sputtering. To ensure consistent wall mapping with EFIT reconstructions, the antenna tilt corresponding to \(\theta > -0.67\)~rad (see Figure~\ref{fig:efitgap}a–b) is excluded from the simulation domain. The simplified 3D geometry used in the calculations, including the walls surrounding the WSHA region, is shown in Figure~\ref{fig:geometry}b.

\begin{figure*}
\centering
\includegraphics[width=\textwidth]{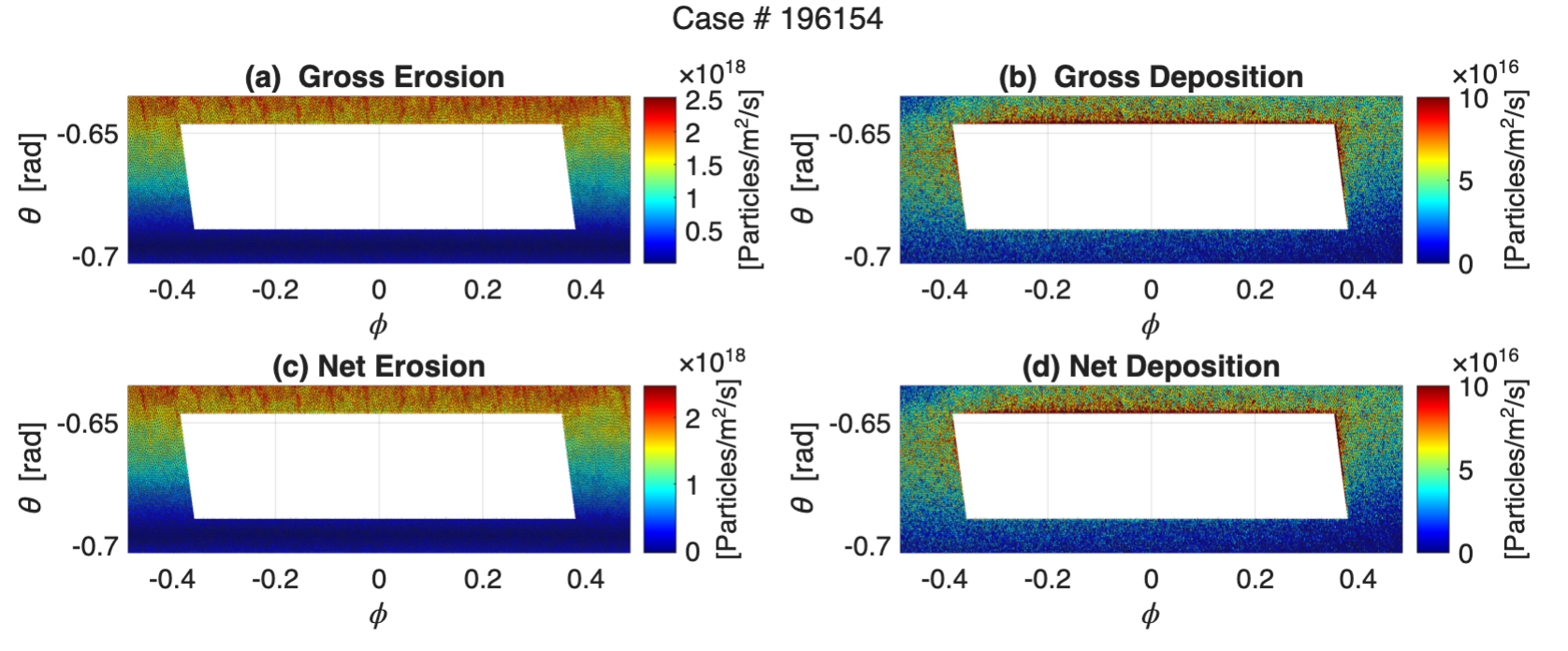}
\caption{Spatially resolved carbon erosion and deposition results from GITRm simulations for DIII-D discharge \#196154. Subfigures show (a) gross erosion, (b) gross deposition, (c) net erosion, and (d) net deposition. Erosion is concentrated in the mid-to-upper poloidal region of the WSHA ($-0.67 < \theta < -0.64$~rad), while deposition is weak and broadly distributed. Net erosion closely mirrors the gross erosion pattern, with re-deposition accounting for less than 5\% of the total eroded carbon. This indicates that most impurity ions escape the local erosion zone and may contribute to SOL and core contamination.}
\label{fig:net_ero_196154}
\end{figure*}

In the GITRm simulations, \(5 \times 10^6\) impurity test particles are initialized based on local erosion fluxes computed from equation~\ref{Eq:grossErosion} and are shown in Figures~\ref{fig:total_gross_196154} and \ref{fig:total_gross_200882}, with initial velocities sampled from a Thomson energy distribution. These particles are tracked until convergence is achieved, typically after \(\sim 5 \times 10^6\) iterations with a time step of \(\Delta t = 1 \times 10^{-9}\)~s.
Figures~\ref{fig:net_ero_196154}a-d show the spatially resolved gross erosion, gross deposition, net erosion, and net deposition for discharge~\#196154. Here  gross erosion is defined as the sum of plasma ($D^+$)- and C-impurity-driven erosion, including self-sputtering effects; gross deposition as the total redeposited particle flux to the surface; net erosion as gross erosion minus gross deposition; and net deposition as gross deposition minus the raw erosion component. Gross erosion is concentrated in the mid-to-upper poloidal region of the WSHA ($-0.67 < \theta < -0.64$~rad), consistent with areas of strong RF sheath potential and high ion flux. Gross deposition is more spatially diffuse and significantly weaker. Net erosion closely follows the gross erosion distribution, while net deposition is weak and broadly spread without strong localization. This pattern is consistent with the relatively large antenna–plasma gap in this discharge ($\sim$7~cm), particularly in the lower WSHA region, where lower plasma density leads to reduced collisionality. Under these low-collisionality conditions, impurity ions have longer mean free paths, reducing the probability of scattering or slowing down near the wall and thereby decreasing the chance of local re-capture.

\begin{figure*}
\centering
\includegraphics[width=\textwidth]{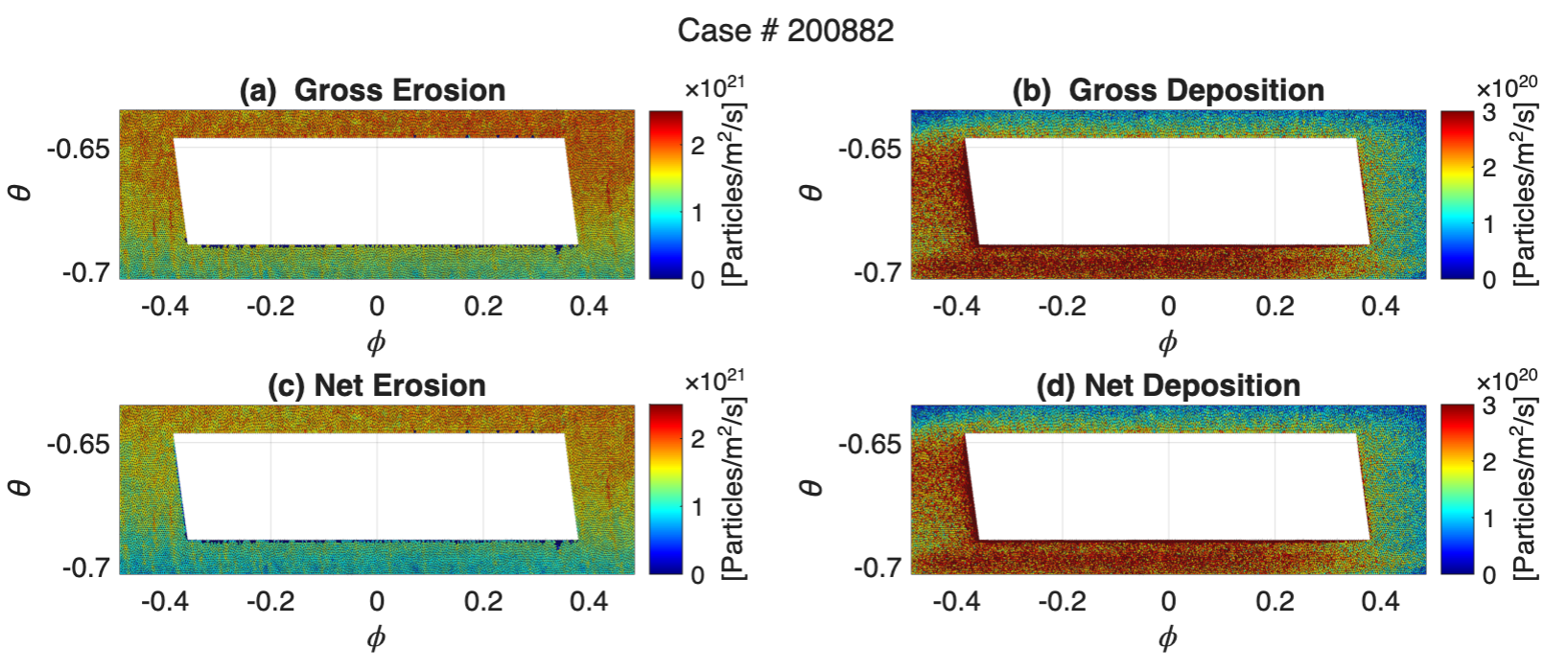}
\caption{Spatially resolved carbon erosion and deposition results from GITRm simulations for DIII-D discharge \#200882. Subfigures show (a) gross erosion, (b) gross deposition, (c) net erosion, and (d) net deposition. Gross and net erosion remain strongest in the mid-to-upper WSHA region, similar to discharge \#196154. However, re-deposition is notably enhanced ($\sim$12\%) in the lower poloidal section, with a larger fraction of eroded material being re-captured locally. }
\label{fig:net_ero_200882}
\end{figure*}

In contrast, discharge~\#200882 (Figure~\ref{fig:net_ero_200882}) exhibits a similar net erosion profile but a markedly different re-deposition behavior. While gross and net erosion again peak in the mid-to-upper WSHA region, net deposition is more localized and pronounced near the lower portion of the antenna. This trend corresponds to the reduced antenna–plasma gap in this discharge ($\sim$4~cm), which results in higher plasma density and stronger collisionality near the lower antenna surfaces. The shorter impurity ion mean free paths in this region increase the likelihood of local re-deposition through scattering and deceleration. This enhanced re-capture demonstrates that impurity retention is strongly modulated not only by geometry but also by local transport conditions.

Overall, the results suggest a clear correlation between local plasma density, collisionality, and impurity re-deposition behavior. In the lower-density, low-collisionality case (\#196154), re-deposition is weak and distributed, with most sputtered carbon escaping the WSHA. In the higher-density, high-collisionality case (\#200882), re-deposition is stronger and spatially concentrated near the lower part of the antenna. Some deposition zones may additionally correspond to field-line-connected, shadowed regions where impurity ions with low parallel velocities or grazing incidence angles ($\sim 80^\circ - 90^\circ$ for the DIII-D helicon antenna as shown later in Figure~\ref{fig:angle_bfield}) intersect nearby surfaces after scattering. However, detailed trajectory-based analysis would be required to confirm such mechanisms.


\subsection{Integrated Carbon Fluxes and Re-deposition Summary}
\label{subsec:44}

To quantify the global balance of carbon erosion and re-deposition, we analyze integrated fluxes from GITRm simulations for discharges~\#196154 and~\#200882. Table~\ref{tab:redeposition_summary} summarizes the gross erosion, gross deposition, net erosion, and net re-deposition rates, along with the corresponding re-deposition fractions.

\begin{table*}
\caption{Area-integrated carbon erosion and re-deposition fluxes [particles/s] from GITRm simulations for the DIII-D helicon antenna.}
\label{tab:redeposition_summary}
\centering
\begin{tabular}{lcccccc}
\hline \hline
\textbf{Discharge} & \textbf{Gross Erosion} & \textbf{Gross Deposition} & \textbf{Net Erosion} & \textbf{Net Re-deposition} & \textbf{Re-deposition Fraction} \\
\hline
\#196154 & \(4.09 \times 10^{17}\) & \(1.63 \times 10^{16}\) & \(3.93 \times 10^{17}\) & \(1.62 \times 10^{16}\) & 3.97\% \\
\#200882 & \(7.20 \times 10^{20}\) & \(9.11 \times 10^{19}\) & \(6.29 \times 10^{20}\) & \(8.84 \times 10^{19}\) & 12.27\% \\
\hline
\end{tabular}
\end{table*}

Discharge~\#200882, characterized by a reduced antenna–plasma gap and elevated local plasma density, exhibits a significantly higher re-deposition fraction of 12.27\%, compared to only 3.97\% in the large-gap case of discharge~\#196154. This increase is consistent with enhanced local collisionality and plasma exposure, which shorten the mean free path of sputtered carbon ions and increase their likelihood of returning to nearby surfaces—particularly in the lower WSHA region. Despite these differences in re-deposition, both discharges exhibit substantial net erosion rates, confirming that the helicon antenna functions as a net impurity source under RF sheath conditions in both coupling regimes.

\subsection{Whole-Device Impurity Transport and Core Penetration}
\label{subsec:45}

\begin{figure}
\centering
\includegraphics[width=\linewidth]{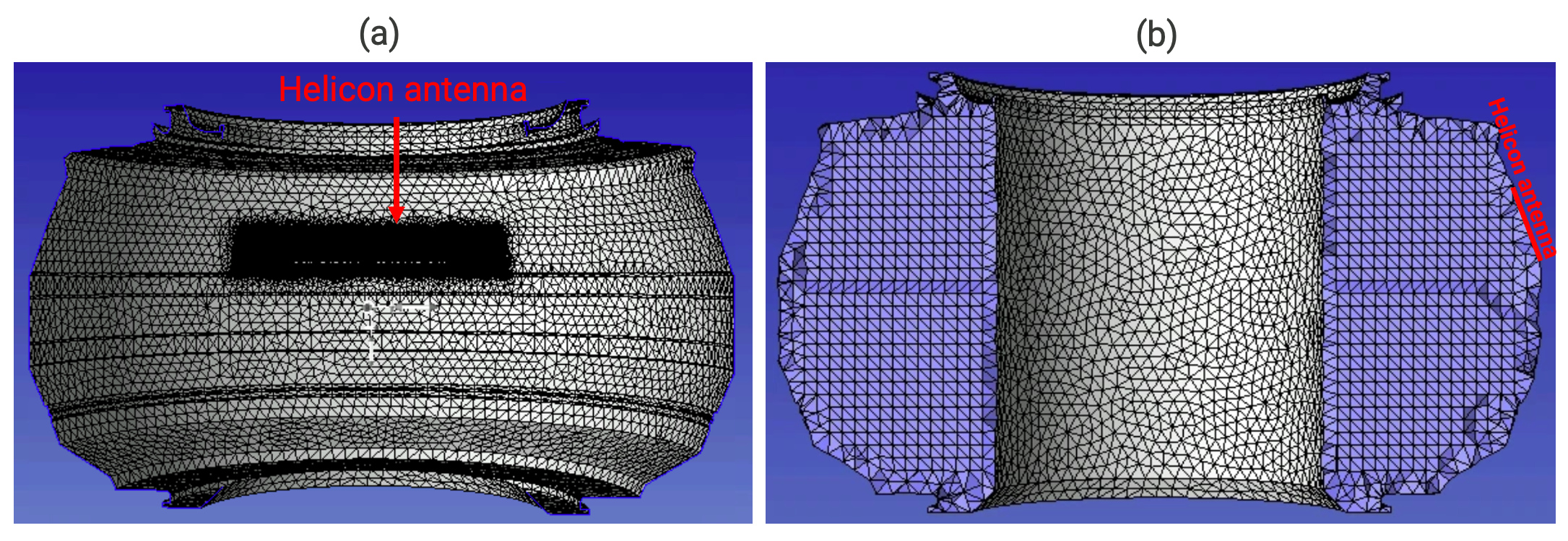}
\caption{Graded unstructured volumetric mesh used in GITRm simulations of whole-device impurity transport in DIII-D. (a) Full-vessel toroidal view showing global mesh coverage. (b) Poloidal cross-sectional view with the helicon antenna and WSHA region (annotated in red).}
\label{fig:gitrm_geometry}
\end{figure}

To evaluate the full impact of helicon-driven erosion on global impurity dynamics, GITRm simulations were performed to compute steady-state 3D carbon density distributions for both discharges using a whole-device unstructured mesh (Figure~\ref{fig:gitrm_geometry}), with mesh refinement around the WSHA region.

\begin{figure}
\centering
\includegraphics[width=\linewidth]{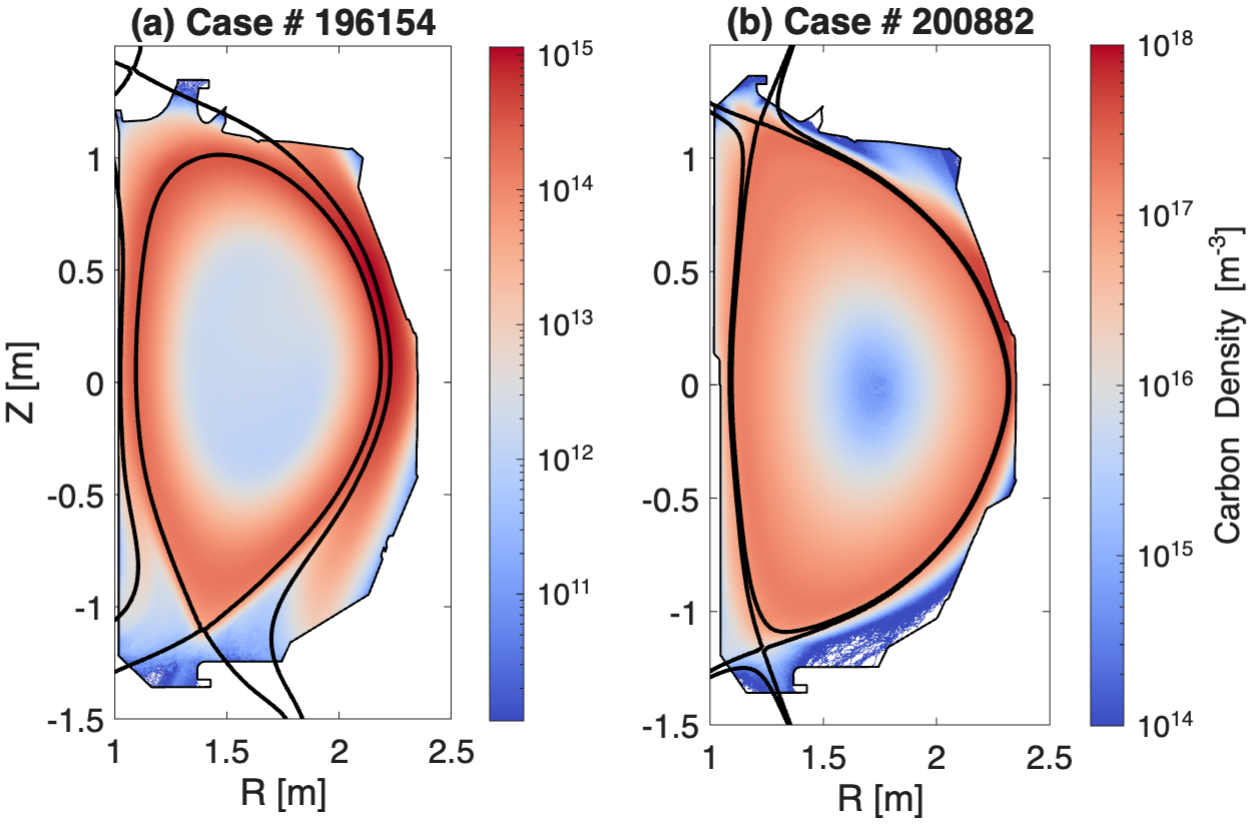}
\caption{Spatial distribution of eroded $n_C$ from GITRm simulations for (a) discharge~\#196154 and (b) discharge~\#200882. Compared to \#196154, discharge~\#200882 exhibits significantly higher impurity content and stronger core penetration, consistent with the reduced antenna–plasma gap and increased local plasma density.}
\label{fig:carbon_density}
\end{figure}

Figure~\ref{fig:carbon_density} presents poloidal cross-sections of the resulting carbon density. In discharge~\#196154, impurity penetration into the confined plasma is modest: the total carbon inventory is \(3.07 \times 10^{15}\) particles, with 34.87\% residing within \(\psi_N \leq 1\), corresponding to an average in-core carbon density of \(5.65 \times 10^{13}\,\mathrm{m}^{-3}\). Helicon-sourced carbon remains localized near the antenna, primarily populating open field lines in the far-SOL, and does not reach the divertor. This indicates limited parallel transport and weak coupling to the confined plasma.

In contrast, discharge~\#200882 exhibits stronger core penetration and broader impurity spread along open field lines. The total carbon inventory increases to \(9.77 \times 10^{17}\), with 58.26\% located inside \(\psi_N \leq 1\), and an average in-core density of \(2.55 \times 10^{17}\,\mathrm{m}^{-3}\). Helicon-sourced carbon in this case reaches divertor-connected flux surfaces and shows enhanced confinement, attributed to stronger RF sheath potentials, tighter plasma coupling, and higher collisionality.

\begin{figure}
\centering
\includegraphics[width=\linewidth]{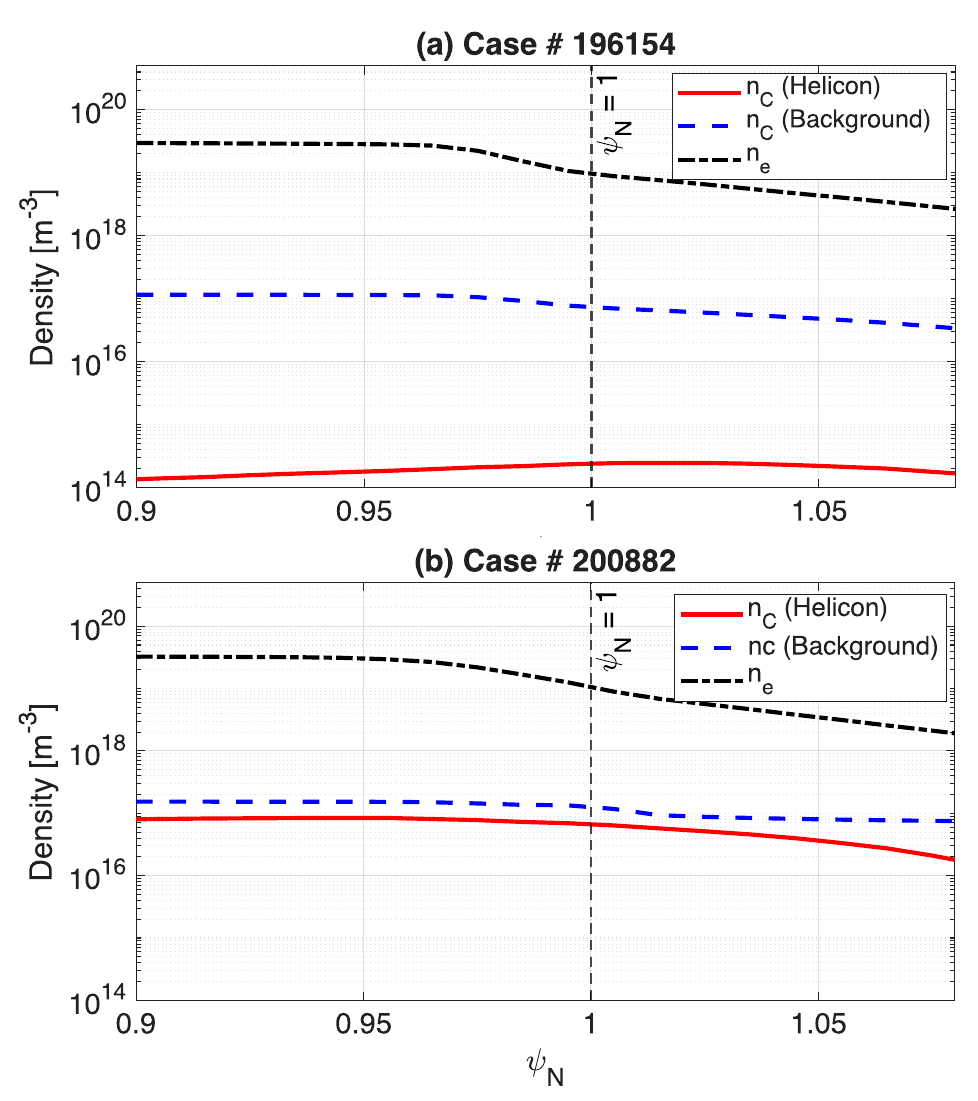}
\caption{Radial profiles of helicon-sourced carbon density \(n_C\) (red solid), background carbon density (blue dashed), and electron density \(n_e\) (black dash-dot) at the OMP from GITRm simulations. (a) In discharge~\#196154, helicon-sourced \(n_C\) remains more than two orders of magnitude lower than in (b) discharge~\#200882, both within the core and the SOL. Overall carbon content in DIII-D remains largely unaffected by the helicon-driven source in both scenarios. }
\label{fig:omp_densities}
\end{figure}

To quantify impurity content at the outer midplane (OMP), Figure~\ref{fig:omp_densities} shows radial profiles of \(n_C\) (helicon-sourced and background) and \(n_e\). In \#196154, helicon-sourced \(n_C\) remains over two orders of magnitude below that in \#200882 across the entire profile. Inside the separatrix, \(n_C^{\mathrm{helicon}} < 10^{14}\,\mathrm{m}^{-3}\) in \#196154, while exceeding \(5 \times 10^{16}\,\mathrm{m}^{-3}\) in \#200882. Moreover, in \#196154, \(n_C^{\mathrm{helicon}}\) is negligible compared to background impurity levels throughout the core and SOL. In contrast, in \#200882, helicon-sourced carbon approaches or surpasses the background near the separatrix ($\psi_N = 1$), indicating stronger antenna-plasma coupling. It should be noted that SOLPS likely underestimates background carbon levels, as it does not contain any plasma interaction with the first wall meaning any main-chamber recycling and sputtering processes will be neglected. Therefore, even in \#200882, the relative importance of \(n_C^{\mathrm{helicon}}\) may be overestimated.

The predictions from GITRm simulations are consistent with the experimental observations shown in Figure~\ref{fig:carbonprofile}, which display no measurable increase in global carbon impurity content during helicon operation in either discharge. In panels (b) and (d), the line-integrated carbon emission remains relatively steady despite the injection of 150–250 kW of helicon power (green trace in (a) and (c)). This agreement supports the conclusion that helicon-sourced carbon does not significantly accumulate in the core plasma under the conditions explored in these discharges.

Overall, these results suggest that in the large-gap, low-collisionality case (\#196154), helicon-induced erosion has minimal impact on the global impurity balance. In the small-gap case (\#200882), helicon-sourced carbon becomes comparable to background impurity levels but does not significantly increase the total impurity content-consistent with experimental observations from DIII-D, which have not reported elevated core carbon levels during helicon operation. However, under divertor-detached conditions-where impurity fluxes from the divertor are suppressed--main-chamber sources such as the helicon antenna may become dominant contributors to the core impurity content. This risk becomes more pronounced with future high-$Z$ wall materials, where even modest impurity influxes can lead to enhanced radiative losses and confinement degradation. This behavior parallels findings from ICRH experiments in WEST, where main-chamber erosion dominated the core impurity inventory under suppressed divertor conditions~\cite{colas:2023, kumar_rfppc_2025}. These results highlight that impurity transport is highly sensitive to local coupling conditions and magnetic geometry. Further experimental diagnostics--including spectroscopy, bolometry, and spectrally resolved imaging--will be essential to distinguish impurity sources from the main chamber and divertor, and to validate impurity transport models in helicon-heated plasmas in DIII-D.

\section{Discussion}
\label{sec:6}

The modeling and analysis presented in this study provide new insights into the coupling between helicon RF power, PMI, and global impurity transport during high-power helicon operation in DIII-D. This section discusses key physical mechanisms, simulation limitations, and their implications for antenna design and impurity control.

\subsection{Sheath Potential Formation and Its Validity}

Rectified DC sheath potentials ($V_{\mathrm{sheath}}$) computed using the COMSOL dielectric-layer method reach several kilovolts in both discharges, with strong localization near the antenna aperture. As shown in Figure~\ref{fig:VDC}, the poloidal distribution of $V_{\mathrm{sheath}}$ follows the geometric modulation imposed by the Faraday screen  with maxima consistently observed in the lower region of the WSHA.  Additional spatial structure along the toroidal direction arises from the discrete antenna modules. This localization arises from a combination of RF field distribution and local plasma conditions--specifically, reduced electron density near the bottom of the antenna, which enhances the rectification process. Discharge~\#196154 exhibits sharper localization at the bottom, while discharge~\#200882, which features broader plasma contact, shows a more spatially extended sheath potential profile, though still bottom-peaked.

\begin{figure}
\centering
\includegraphics[width=\linewidth]{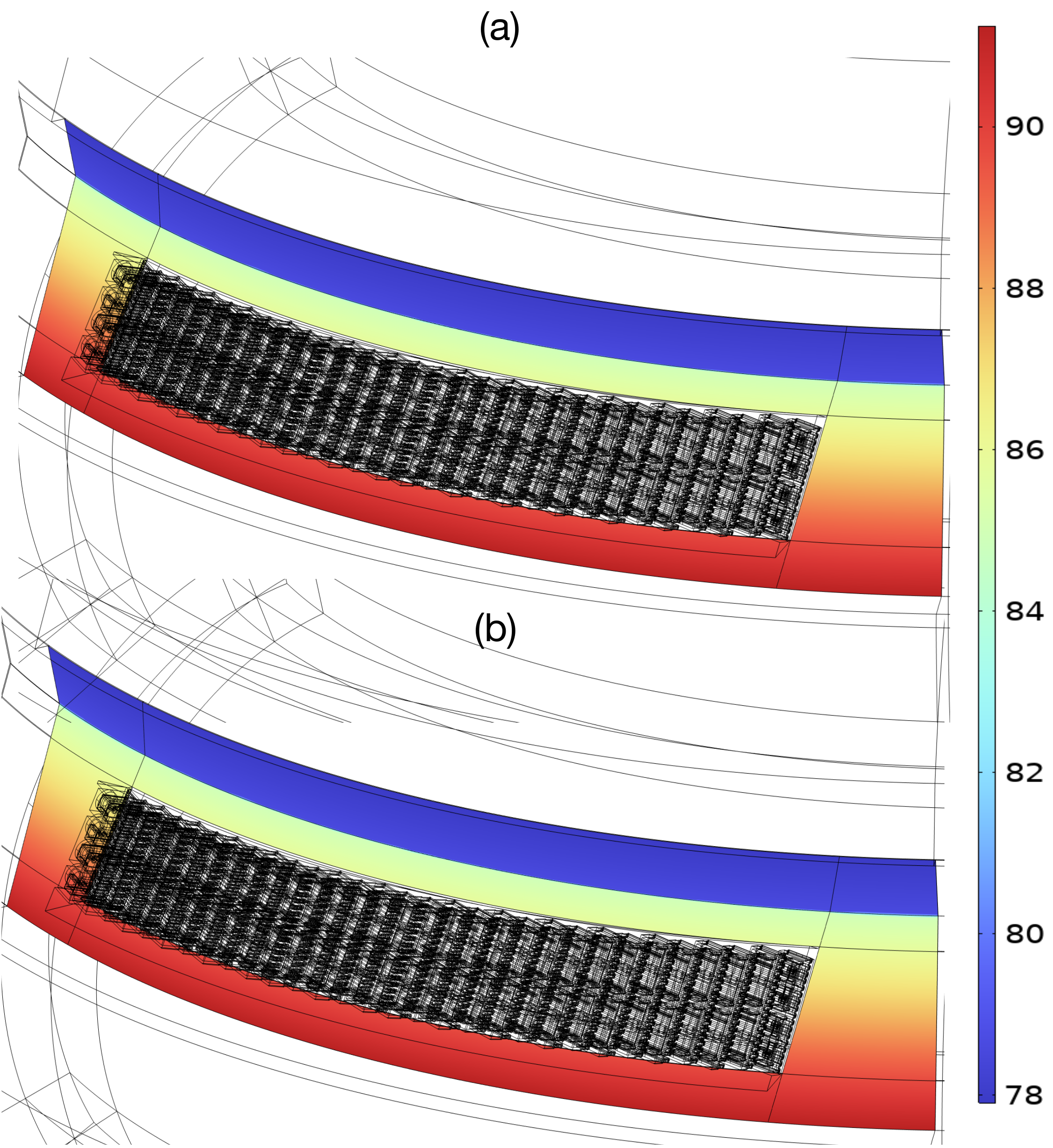}
\caption{Magnetic field incidence angle relative to the antenna surface for (a) discharge~\#196154 and (b) discharge~\#200882. In both cases, the helicon antenna is oriented nearly tangential to the magnetic field, with incidence angles close to $90^\circ$, especially along the lower WSHA. This grazing-angle configuration is challenging for RF sheath modeling, as most existing boundary conditions are not validated for such extreme alignment.}
\label{fig:angle_bfield}
\end{figure}

A key geometric factor contributing to this behavior is the nearly tangential orientation of the sheath surfaces with respect to the confining magnetic field. As shown in Figure~\ref{fig:angle_bfield}, the magnetic field incidence angle approaches $90^\circ$ along much of the WSHA surface, particularly near the bottom. This extreme grazing-angle configuration presents a substantial challenge for RF sheath modeling. Most existing sheath boundary conditions--whether derived from surface impedance models, dielectric-layer formulations, or kinetic sheath theory--are developed for perpendicular or moderately oblique magnetic field incidence and are not well validated for tangential configurations\cite{myra2021tutorial}. Consequently, the absolute magnitude and localization of $V_{\mathrm{sheath}}$ under these conditions remain uncertain.

Despite these limitations, the qualitative trends in $V_{\mathrm{sheath}}$ predicted by COMSOL are consistent with physical expectations. Sheath rectification strength increases with decreasing electron density, and the Faraday screen's periodicity drives poloidal modulation. However, accurate modeling of sheath potentials in helicon systems with grazing field-line geometry will require further development of boundary conditions tailored for near-parallel magnetic field alignment, as well as targeted experimental validation in relevant regimes.

\subsection{Slow Wave Propagation and Modeling Limitations}
\label{sec:6.2}

Accurate modeling of slow wave dynamics near the helicon antenna is critical for predicting sheath excitation and impurity sourcing. Analytical solutions in the cold plasma collisional regime~\cite{tierens2024slow} show that slow waves become propagative in a narrow edge layer and form highly localized resonance cone structures. These waves are characterized by short perpendicular wavelengths (\(\lambda_\perp \lesssim 1\)~mm) and steep field gradients near the antenna aperture. However, resolving these fine-scale structures remains a major challenge in full-wave 3D simulations. As demonstrated in the reference \citen{Wouter_2024}, even with third-order finite elements, accurate resolution of the RF electric field near resonance cone tangency points requires over \(10^3\) mesh elements per local radius of curvature. Standard COMSOL simulations are not sufficient to resolve these fields, leading to a systematic underestimation of electric field amplitudes and, by extension, rectified sheath potentials.

Moreover, current models treat the plasma as a cold dielectric and omit kinetic mechanisms such as Landau damping, finite Larmor radius effects, and velocity space resonances. These effects may influence slow wave damping, spectral broadening, and field localization near material surfaces.

Altogether, the combination of limited mesh resolution and the absence of kinetic physics likely leads to an underprediction of both the magnitude and localization of \(V_{\mathrm{sheath}}\) in our simulations. Future improvements will require hybrid modeling approaches that couple full-wave solvers to kinetic or reduced-order sheath models capable of capturing slow wave physics under grazing magnetic field alignment and strong collisional damping.

\subsection{Coupling Condition and Antenna–Plasma Gap}

The antenna–plasma gap plays a critical role in determining both RF power coupling efficiency and the spatial extent of RF sheath formation. In discharge~\#200882, the reduced gap of approximately 4~cm brings the antenna and surrounding walls into closer contact with the edge plasma. This results in stronger local collisionality, enhanced re-deposition near the lower WSHA, and improved coupling efficiency. In contrast, the larger 7~cm gap in discharge~\#196154 reduces plasma exposure, weakens the sheath-connected surface area, and limits local impurity recapture.

These results confirm that tighter coupling conditions--while beneficial for RF performance--also intensify PMI. Operational trade-offs between coupling efficiency and impurity control must therefore be considered in antenna design and discharge optimization.

\subsection{Sputtering Contributions from D$^+$ Ions}

While carbon self-sputtering dominates total erosion under RF sheath conditions, simulations show that D$^+$ ions contribute a non-negligible fraction to material erosion at the WSHA. In both discharges, D$^+$-induced erosion accounts for approximately 1\% of the total gross carbon erosion flux. This contribution arises from RF sheath acceleration, which increases the impact energy of D$^+$ ions into the 1–5~keV range--where sputtering yields become appreciable, particularly at tangential incidence angles (see Figures~\ref{fig:rustbca_d_on_c}, \ref{fig:total_gross_196154}, and \ref{fig:total_gross_200882}). The contrast between D$^+$-driven and impurity-driven erosion further suggests that reducing background impurity levels--such as through improved divertor conditions or wall conditioning--could significantly lower overall erosion rates. This effect is expected to become even more relevant with planned high-$Z$ wall upgrades in DIII-D.

Although \( Y_{\mathrm{eff}}(D^+)\) on carbon remains relatively low--typically peaking below 0.06 atoms/ion--the combination of high incident flux and sheath-driven energy enhancement leads to measurable erosion across exposed antenna surfaces. This finding has important implications for future wall materials. If graphite tiles at the WSHA are replaced with high-$Z$ metals (e.g., tungsten), D$^+$ sputtering can become more consequential. At kilovolt-level sheath potentials, the sputtering yield of D$^+$ on tungsten falls in a similar range as that for carbon or oxygen. Moreover, even modest tungsten erosion can introduce high-$Z$ impurities into the plasma, leading to increased core radiative losses and confinement degradation. Therefore, accurate quantification of D$^+$-driven erosion is essential for future helicon antenna designs and operational scenarios involving high-$Z$ first wall materials.

\subsection{Sensitivity to Main Ion Flow Assumptions}
\label{sec:discussion2}

The evolution and spatial distribution of carbon charge states in the SOL and near-antenna region are strongly influenced by the assumed parallel flow of the main ions. As discussed in reference~\citen{schmid_iter_w_erosion}, both the magnitude and direction of this flow critically affect impurity retention, re-ionization, and wall loading. In cases with strong parallel flow (e.g., \(M \gtrsim 0.5\)) and weak cross-field transport, impurities are rapidly convected toward the divertor, reducing their ionization progression and limiting exposure to the main chamber walls like helicon antenna region. In contrast, in stagnant plasmas with negligible parallel flow and enhanced perpendicular transport, high charge state impurities can migrate from the separatrix all the way to the wall. There, they are accelerated by RF-sheath electric fields, resulting in enhanced sputtering and increased local erosion.

This sensitivity is particularly important in the context of the DIII-D helicon antenna, where magnetic field lines are long and the antenna surface is nearly tangential to the local field. In such configurations, small variations in parallel flow can significantly affect impurity trajectories and re-ionization histories. The current SOLPS-based modeling relies on representative, but experimentally unvalidated, parallel flow profiles. As a result, predictions of charge state distributions and impurity transport near the antenna should be interpreted with caution. Direct measurements of main ion flow using diagnostics such as Mach probes or Doppler spectroscopy would provide critical constraints and substantially improve the reliability of impurity transport modeling in RF-heated edge plasmas.

\subsection{Implications for Future Helicon Operation}

The inability to fully resolve short-wavelength, sheath-driving fields in current full-wave simulations underscores the need for sheath-aware operational strategies in future high-power helicon scenarios. While reducing the antenna–plasma gap enhances RF coupling efficiency, it also increases the risk of localized surface erosion due to elevated rectified sheath potentials. Operational regimes must therefore be optimized to balance efficient wave coupling with impurity control and long-term PFC integrity.

Potential mitigation approaches include tailoring discharge parameters—such as edge density, magnetic configuration, and heating power—to minimize RF sheath rectification in high-field regions.  Real-time monitoring of impurity content and radiative losses, combined with predictive impurity modeling, will be essential for maintaining plasma purity and avoiding core contamination—especially in devices with high-$Z$ wall materials. To support future operational planning, improved modeling frameworks capable of capturing multi-scale wave-sheath interactions under grazing-angle magnetic fields will be critical. 
\section{Summary}
\label{sec:7}

This study applied the STRIPE (Simulated Transport of RF Impurity Production and Emission) framework to model carbon erosion and impurity transport near the high-power helicon antenna in the DIII-D tokamak. STRIPE integrates background plasma profiles from SOLPS, rectified sheath potential calculations from COMSOL, angle- and energy-resolved sputtering yields from RustBCA, and full-orbit impurity transport using GITR and GITRm. The framework was applied to two H-mode discharges--\#196154 and \#200882--with antenna–plasma gaps of approximately 7~cm and 4~cm, respectively, to assess the impact of RF sheath physics and geometric coupling on plasma–surface interactions.

COMSOL simulations predict rectified sheath potentials ($V_{\mathrm{sheath}}$) exceeding 2–3~kV in both discharges, with strong localization near the bottom of the WSHA region, where the antenna surfaces are nearly tangential to the magnetic field. This grazing-angle geometry poses a significant challenge for existing sheath models and introduces uncertainty in the predicted sheath amplitudes. Nevertheless, the simulated $V_{\mathrm{sheath}}$ distribution correlates well with regions of peak ion flux and carbon erosion, supporting the validity of the dielectric-layer approach for this application.

Erosion in both discharges is primarily driven by carbon self-sputtering, with effective yields ($Y_\mathrm{eff}$) reaching up to 7.5–8 atoms/ion in the small-gap case. Although D$^+$-induced sputtering yields remain below 0.06 atoms/ion, RF sheath acceleration to multi-keV energies enables D$^+$ ions to contribute approximately 1\% of the total erosion flux. While this contribution is minor for graphite walls, it becomes more consequential in future high-$Z$ wall scenarios (e.g., tungsten), where even modest D$^+$-driven erosion can introduce radiative high-$Z$ impurities into the core. These findings also highlight that maintaining a cleaner plasma--by minimizing impurity concentrations--can significantly reduce overall erosion, leaving D$^+$ as the dominant but less damaging erosion driver.

GITRm simulations show total gross erosion reaching $7.2 \times 10^{20}$~particles/s in discharge~\#200882, compared to $4.1 \times 10^{17}$~particles/s in \#196154. The corresponding net re-deposition fractions are approximately 12.3\% and 4.0\%, respectively. These differences are attributed to higher collisionality and increased plasma contact in the small-gap case. Carbon transport analysis further reveals that 58\% of eroded carbon penetrates the confined plasma in discharge~\#200882, compared to only 35\% in discharge~\#196154.

Taken together, these results emphasize the importance of self-consistent sheath modeling and species-resolved sputtering analysis when evaluating helicon antenna performance. Accurate prediction of $V_{\mathrm{sheath}}$ under grazing-angle magnetic fields remains a key challenge due to the limitations of existing sheath boundary models. Future work should focus on improved treatment of near-parallel sheath angles, kinetic damping mechanisms, and short-wavelength slow wave structures.

Finally, the simulation results are consistent with experimental observations from DIII-D, which have not shown elevated core impurity levels during helicon operation in the present graphite-wall configuration. However, under reactor-relevant conditions involving high-power operation and high-$Z$ PFCs, even modest impurity sourcing from main-chamber structures—such as the helicon antenna—could lead to significant core contamination due to enhanced radiative losses from high-$Z$ species. This emphasizes the importance of continued experimental diagnostics and advanced impurity transport modeling to evaluate the role of RF-driven erosion and main-chamber impurity sources in future high-performance helicon scenarios in DIII-D and other devices.
\section*{Disclaimer}
 This report was prepared as an account of work sponsored by an agency of the United States Government. Neither the United States Government nor any agency thereof, nor any of their employees, makes any warranty, express or implied, or assumes any legal liability or responsibility for the accuracy, completeness, or usefulness of any information, apparatus, product, or process disclosed, or represents that its use would not infringe privately owned rights. Reference herein to any specific commercial product, process, or service by trade name, trademark, manufacturer, or otherwise does not necessarily constitute or imply its endorsement, recommendation, or favoring by the United States Government or any agency thereof. The views and opinions of authors expressed herein do not necessarily state or reflect those of the United States Government or any agency thereof.

\section*{Acknowledgement}
 This research used resources of the Fusion Energy Division, FFESD and the ORNL Research Cloud Infrastructure at the Oak Ridge National Laboratory, which is supported by the Office of Science of the U.S. Department of Energy under Contract No. DE-AC05-00OR22725. This material is also based upon work supported by the U.S. Department of Energy, Office of Science, Office of Fusion Energy Sciences, using the DIII-D National Fusion Facility, a DOE Office of Science user facility, under Award(s) DE-FC02-04ER54698. RPI work is supported by the US DOE under the contracts DE-SC0024369 and DE-SC0021285.
 \appendix
\section{Global Impurity Transport with GITRm}
\label{app:gitrm}

GITRm~\cite{nath:2023, Nath:2025} is a 3D unstructured-mesh Monte Carlo particle-tracking code developed to simulate impurity transport, PMI, and material erosion in magnetically confined fusion plasmas.  It builds upon the PUMIPic infrastructure\cite{Diamond:2021} and is designed for execution on GPU-accelerated and distributed-memory computing systems. It models particle dynamics under Lorentz forces, collisional drag and diffusion, anomalous transport, and sheath electric fields. The kinetic model is summarized below.

The Boltzmann transport equation in the trace approximation evolves the distribution function \(f_z\):
\begin{eqnarray}
&&
\mathbf{v}_z \cdot \nabla f_z 
\quad
+\quad
\frac{q}{m_z}
\Bigl(
\mathbf{E} + \mathbf{v}_z \times \mathbf{B}
\Bigr)
\cdot
\frac{\partial f_z}{\partial \mathbf{v}_z}
\nonumber\\
&&
\quad
=\quad
C_{zb}\bigl(f_z,f_b\bigr)
\quad+\quad
S_1 \quad+\quad S_2 \quad+\quad S_3.
\end{eqnarray}
Here, \(f_z\) is the impurity distribution function in phase space, \(\mathbf{v}_z\) is the particle velocity, \(\mathbf{r}\) the position vector, \(q\) the particle charge, \(m_z\) the mass, \(\mathbf{E}\) and \(\mathbf{B}\) the electric and magnetic fields, \(C_{zb}\) the collision operator with the Maxwellian background plasma \(f_b\), and \(S_1\), \(S_2\), and \(S_3\) are source terms representing plasma erosion, self-erosion, and ionization or recombination.

The Lorentz force acting on each particle is given by:
\begin{equation}
\mathbf{F}_{\mathrm{Lorentz}} = q \left(\mathbf{E} + \mathbf{v}\times\mathbf{B}\right),
\end{equation}
where \(\mathbf{v}\) is the instantaneous particle velocity.

The collisional drag force describing slowing down relative to the plasma is:
\begin{equation}
\mathbf{F}_{\mathrm{Drag}} = -\nu_s\, m\, \mathbf{v},
\end{equation}
where \(\nu_s\) is the slowing-down frequency and \(m\) is the particle mass.

Pitch-angle and energy diffusion are modeled via:
\begin{equation}
\frac{d}{dt}\left|\mathbf{v}-\bar{\mathbf{v}}\right|^2_{\parallel} = \nu_{\parallel}\, v^2,
\end{equation}
\begin{equation}
\frac{d}{dt}\left|\mathbf{v}-\bar{\mathbf{v}}\right|^2_{\perp} = \nu_{\perp}\, v^2,
\end{equation}
where \(\bar{\mathbf{v}}\) is the mean plasma velocity, \(\nu_{\parallel}\) and \(\nu_{\perp}\) are the collisional frequencies for parallel and perpendicular scattering, and \(v=|\mathbf{v}|\).

The kinetic energy decay due to collisional energy loss is expressed as:
\begin{equation}
\frac{d\epsilon}{dt} = -\nu_\epsilon\, \epsilon,
\end{equation}
where \(\epsilon = \frac{1}{2}\,m\,v^2\) is the kinetic energy and \(\nu_\epsilon\) is the energy relaxation frequency.

The velocity update per timestep combines deterministic slowing down and stochastic diffusion:
\begin{eqnarray}
\mathbf{v} &=& 
\mathbf{v}_0\left(1 - \frac{\nu_\epsilon \Delta t}{2}\right)
\Bigl[
\hat{e}_1\Bigl(1 + \eta_1\sqrt{\nu_{\parallel}\Delta t}\Bigr)
\nonumber\\
&&
\quad
+\,
\bigl|\eta_2\bigr|\sqrt{\frac{\nu_{\perp}\Delta t}{2}}
\hat{e}_2\cos(2\pi\xi)
\nonumber\\
&&
\quad
+\,
\bigl|\eta_2\bigr|\sqrt{\frac{\nu_{\perp}\Delta t}{2}}
\hat{e}_3\sin(2\pi\xi)
\Bigr]
\nonumber\\
&&
\quad
-\,\nu_s \Delta t\,\mathbf{v}_0.
\end{eqnarray}
In this equation, \(\mathbf{v}_0\) is the velocity before the update, \(\hat{e}_1\) is the unit vector parallel to \(\mathbf{v}_0\), and \(\hat{e}_2\) and \(\hat{e}_3\) are orthonormal vectors spanning the perpendicular plane. The variables \(\eta_1\) and \(\eta_2\) are independent standard normal random variables, and \(\xi\) is uniformly distributed in \([0,1]\).

Anomalous cross-field diffusion is applied as:
\begin{equation}
\mathbf{r} = \mathbf{r}_0 + \sqrt{4\,D\,\Delta t}\,\hat{b}_\perp,
\end{equation}
where \(\mathbf{r}_0\) is the particle position before diffusion, \(D\) is the anomalous diffusion coefficient, and \(\hat{b}_\perp\) is a random unit vector perpendicular to the local magnetic field.

Ionization and recombination reactions are modeled by sampling a reaction probability in each timestep. The mean time between reactions is:
\begin{equation}
\tau_{\mathrm{atm}} = \frac{1}{\langle \sigma v \rangle\,n},
\end{equation}
where \(\langle \sigma v \rangle\) is the Maxwellian-averaged reaction rate coefficient and \(n\) is the plasma density. The probability of a reaction during timestep \(\Delta t\) is:
\begin{equation}
P = 1 - \exp\left(-\frac{\Delta t}{\tau_{\mathrm{atm}}}\right).
\end{equation}

The unstructured mesh in this study is generated specifically for the 3D geometry surrounding the DIII-D helicon antenna, including PFCs such as the antenna front surfaces and adjacent wall. Anisotropic refinement is applied in regions of strong plasma gradients and near sheath-relevant surfaces. Plasma fields and collision frequencies are interpolated to particle positions using linear finite-element shape functions on this mesh. Surface tagging is employed to identify antenna-facing facets for applying sheath electric fields and tracking erosion and re-deposition patterns.

Further details on solver algorithms, parallel performance, and validation studies are provided in reference \citen{nath:2023}.

\section{DIII-D Helium Beam Diagnostics}
\label{app:heData}

The thermal helium beam diagnostic at DIII-D was designed to measure the plasma parameters in front of the helicon antenna. Installed at the $195^{\circ}$ midplane port, the system consists of a PEV-1 piezo-valve connected to a helium gas line maintained at a pressure of $300$–$450~\mathrm{mTorr}$ by a pressure regulator. The gas is delivered from the piezo-valve through a $3.175~\mathrm{mm}$-diameter duct to a puff location situated below the helicon antenna.

An in-vessel mirror was installed to collect light emitted from the plasma along a radial line-of-sight that intersects the helium cloud in front of the puffing region. The collected light is focused by a lens and coupled into a fiber bundle consisting of 18 fibers (CeramOptec WVNSS, $300~\mu\mathrm{m}$ core diameter, $\mathrm{NA}=0.12$). An identical lens and fiber bundle are installed side-by-side; both are aligned to view nearly the same plasma region, separated by only a few millimeters.

Light from one fiber bundle is directed to an IsoPlane 320STC spectrometer equipped with a 600~lines/mm grating (1~µm blaze) and an EMCCD camera (Princeton Instruments ProEM-HS 1024BX3, $1024\times1024$ pixels). This spectrometer is used to measure the neutral helium lines at 667.8, 706.5 and 728.1~nm. The second fiber bundle is routed to an IsoPlane 160 spectrometer (600~lines/mm grating, 500~nm blaze) coupled to the same model of EMCCD camera. This spectrometer is primarily used for impurity surveys and for measurements of visible neutral helium emission. The optical system was absolutely calibrated using a NIST-traceable tungsten–halogen standard lamp.

The lines-of-sight intersect the helium beam almost perpendicularly. The first observation point is located approximately $45~\mathrm{mm}$ from the helicon antenna. The viewing geometry covers a radial extent of about $67~\mathrm{mm}$ with a spatial resolution of $\sim 3.7~\mathrm{mm}$; each individual view corresponds to a spot size of approximately $3~\mathrm{mm}$ in diameter. The first line of sight is about $42~\mathrm{mm}$ away from the center of the helicon antenna.

The inference of the electron density $n_e$ and electron temperature $T_e$ profiles from the measured helium line emission is performed using collisional–radiative (CR) modeling based on the ADAS dataset \texttt{pec96\#he\_pju\#he0.dat}. This file provides the photon emissivity coefficients (PECs) of neutral helium lines under stationary plasma conditions, from which the expected line intensity ratios can be calculated. 
Thus, the local plasma parameters $n_e$ and $T_e$ are obtained by minimizing the difference between the experimentally measured line ratios and the theoretical values predicted by the CR model.

\section*{References}
\bibliographystyle{ieeetr}  
\bibliography{diiid-helicon.bib}
\end{document}